\documentclass[12pt,preprint]{aastex62}
\usepackage{CJK}
\usepackage{xcolor}
\usepackage{multirow}
\begin{document}
\begin{CJK*}{UTF8}{gbsn}
\title{A Phenomenological Model for the Light Curve of three Quiescent Low-inclination Dwarf Novae and one Pre-Cataclysmic Variable}
\author[0000-0002-4280-6630]{Zhibin Dai (戴智斌)}
\affiliation{Yunnan Observatories, Chinese Academy of Sciences, 396 Yangfangwang, Guandu District, Kunming, 650216, P. R. China.}
\affiliation{Key Laboratory for the Structure and Evolution of Celestial Objects, Chinese Academy of Sciences, 396 Yangfangwang, Guandu District, Kunming, 650216, P. R. China.}
\affiliation{Center for Astronomical Mega-Science, Chinese Academy of Sciences, 20A Datun Road, Chaoyang District, Beijing, 100012, P. R. China.}
\affiliation{University of Chinese Academy of Sciences, No.19(A) Yuquan Road, Shijingshan District, Beijing, 100049, P.R.China}
\author[0000-0003-4373-7777]{Paula Szkody}
\affiliation{University of Washington, Seattle, WA, 98195, USA.}
\author[0000-0001-6894-6044]{Mark Kennedy}
\affiliation{Jodrell Bank Centre for Astrophysics, School of Physics and Astronomy, The University of Manchester, Manchester M13 9PL, UK}
\author[0000-0001-7566-9436]{Jie Su (苏杰)}
\affiliation{Yunnan Observatories, Chinese Academy of Sciences, 396 Yangfangwang, Guandu District, Kunming, 650216, P. R. China.}
\affiliation{Key Laboratory for the Structure and Evolution of Celestial Objects, Chinese Academy of Sciences, 396 Yangfangwang, Guandu District, Kunming, 650216, P. R. China.}
\affiliation{Center for Astronomical Mega-Science, Chinese Academy of Sciences, 20A Datun Road, Chaoyang District, Beijing, 100012, P. R. China.}
\author{N. Indika Medagangoda}
\affiliation{Arthur C. Clarke Institute for Modern Technologies, Sri Lanka.}
\author{Edward L. Robinson}
\affiliation{University of Texas at Austin, Austin, TX, , USA.}
\author[0000-0003-4069-2817]{Peter M. Garnavich}
\affiliation{University of Notre Dame, Notre Dame, IN, 46556, USA.}
\author{L. Malith M. De Silva}
\affiliation{Sabaragamuwa University of Sri Lanka.}
\correspondingauthor{Zhibin Dai }
\email{zhibin\_dai@ynao.ac.cn}

\begin{abstract}
We used the light curve code XRBinary to model the quiescent K2 light curves of three low-inclination cataclysmic variables (CVs): 1RXS\,J0632+2536 (J0632+2536), RZ\,Leo, TW\,Vir and the pre-CV WD\,1144+011. Optimized light curve models were obtained using a nonlinear fitting code NMfit and visualized by Phoebe 2.0. The disk model of J0632+2536 shows that one hotspot at the edge of the disk is enough to describe its light curve, while the other two dwarf nova (DN): RZ\,Leo and TW\,Vir require two hotspots. A typical pre-CV model with a weak irradiation effect for WD\,1144+011 can explain its single-hump modulation, and the newly observed spectrum confirms its previous classification. The synthetic analyses for the DN clearly indicate that phase zero of the double-hump modulations occurs around the secondary minimum and the primary hump is mainly caused by the hotspot at the edge of the disk. The quiescent disk has a flat temperature distribution with a power index of $\sim0.11$. The disk model of RZ\,Leo implies a truncated disk, supporting its previously speculated classification as an intermediate polar (IP). Except for the IP model of RZ\,Leo, which lacks a component related to the inferred accretion curtain, the models of J0632+2536, TW\,Vir and WD\,1144+011 are consistent with results from the Gaia mission. The derived masses and radii of the secondaries of the three DN are consistent with the semi-empirical relations for CV donor stars, while their effective temperatures are higher than the predictions. Irradiation of the donor stars is investigated to explain this discrepancy.
\end{abstract}
\keywords{Stars : binaries : close; Stars : cataclysmic variables; Stars : white dwarfs}

\section{Introduction}

Dwarf novae (hereafter DN) are a subtype of primarily non-magnetic cataclysmic variables (hereafter CVs), in which a white dwarf primary accretes matters from a Roche-lobe filling late-type star via the inner (L1) Lagrange point \citep{war03}. An accretion disk can extend to the white dwarf surface via viscous processes between adjacent accretion annuli (e.g., friction and shear) due to a weak magnetic field of the white dwarf in DN (B$<$10$^{6}$\,G). Systems in which the magnetic field of the white dwarf is large enough (10$^{6}<B\leqslant$\,10$^{7}$\,G) such that the accretion disk is disrupted inside the white dwarf magnetosphere and material begins to follow the magnetic field lines are called Intermediate Polars (hereafter IPs). In the following, we assume that the disk outside of the magnetosphere in an IP is equivalent in structure to the disk in a regular non-magnetic DN. The interaction between the ballistic stream leaving the L1 point and the accretion disk forms a region of energy release at the edge of the disk called a hotspot. The combination of an accretion disk and a hotspot (hereafter just called the disk model) has been used as the typical accretion model for CVs \citep[e.g.][]{sma70,war03} and has been quite successful in describing the asymmetrical eclipse profile of several high-inclination CVs during quiescence \citep[e.g.][]{bru96,woo86}. This model can also be used to explain a symmetrical CV eclipse profiles during outburst \citep[e.g.][]{kat03,bak15}, which is caused by the significant flux increase of the disk (i.e., the luminous accretion disk almost overwhelms the relatively faint white dwarf and hotspot).

Several high-inclination CVs in quiescence have been comprehensively studied using this disk model \citep[e.g.][]{coo84,bai91,mca15}. By detecting variations in the mid-eclipse times, substellar objects have been suggested to exist in several DN (e.g., V2051\,Oph \citep{qia15} and EM\,Cyg \citep{dai10a}). In addition, many synthetic light curve analysis methods (e.g., BINSYN program \citep{lin12}, Eclipsing Light Curve Code \citep{oro00} and the cool-disk model \citep{khr11}) have been developed to analyze CV eclipse light curves. The irregular eclipse light curves of quiescent CVs are composed of the occultations of multiple components including the white dwarf, the accretion disk and hotspots \citep[e.g.][]{sma94,fel04,lit14,mca15}. High-inclination CVs only show a single eclipse in one orbit (i.e., the secondary eclipse of the red dwarf is invisible \cite[e.g.][]{krz65,bai88}), since the red dwarf in a CV system is usually regarded to be a very faint component compared with the accretion disk and the white dwarf. This ``single-eclipse" feature implies that CV eclipse light curves cannot reveal full information about the red dwarf companion. Due to the complexity of the quiescent CV eclipse light curves, high time resolution is necessary to decompose all components. However, over 70\% of CVs with short orbital periods ($<$\,3\,hr) are fainter than 17\,mag in quiescence based on the updated CV catalogue (RKcat Edition 7.24 first published in \cite{rk724}). Therefore, the majority of CVs have only low time-resolution photometry comprised of blended flux from many components, which makes modelling these systems difficult. For example, recent discussions concerning two new-found CVs with deep eclipses carried out by \cite{kju15} and \cite{ken16} clearly indicated that the model light curves cannot perfectly fit the observed eclipsing light curves during ingress and outside of eclipse. Assuming a deep eclipse is a common feature of CVs with orbital inclination higher than 80$^{\circ}$, the fraction of low-inclination CVs can be simply estimated to be around 90\%. To determine a general model independent of inclination, it is necessary to consider low-inclination CVs.

The unprecedented light curves from the Kepler K2 mission \citep{how14}, with nearly continuous photometric coverage for 1-3 months at different pointings (Campaigns) along the ecliptic provide an excellent database to study quiescent CV light curves. K2 Campaign 0 (K2-C0) was an engineering test in the early stage of the K2 program and only covered $\sim$\,35 days since the spacecraft was not in fine point during the beginning of the campaign, while K2 Campaign 1 (K2-C1) covered a complete period of 80 days. We focus in this paper on the phased light curves of four systems: 1RXS\,J0632+2536 (hereafter J0632+2536) and TW\,Vir which are both DN, RZ\,Leo which is an unusual IP that also has displayed DN outbursts, making it one of the few systems to be both a magnetic system and a DN, and WD\,1144+011 which is a pre-CV (meaning the secondary is likely not filling its Roche lobe). J0632+2536 was observed in K2-C0, while TW\,Vir, RZ\,Leo and WD\,1144+011 were observed in K2-C1. The phased light curves are investigated in detail using the synthesis methods XRBinary and NMfit described in Section 3. Due to the lack of any eclipse feature, all four objects are likely low-inclination systems. The preset model parameters are discussed in Sections 4 and 5. The details of the white dwarf accretion structure during quiescence and the physical parameters of the stars in each system are further discussed and visualized in Section 6.

\section{Phased Light Curves}

\cite{dai16} used the PyKE suite of software tools developed by the Guest Observer Office \citep{sti12} to extract the K2 light curves of J0632+2536, RZ\,Leo and WD\,1144+011, and derive their orbital periods using traditional period finding techniques (e.g., Lomb-Scargle periodogram \citep{lom76,sca82}, phase dispersion minimization \citep{ste78}) and their corresponding phased light curves. Due to the unstable orbital modulations of TW\,Vir, \cite{dai17} applied a phase-correcting method to obtain its orbital period and the phased data from the quiescent data.  

Based on the continuous K2 data, we attempted to model the mean orbital light curves of each system by using an XRBinary light curve synthesis code developed by E. L. Robinson \footnote{\url{http://www.as.utexas.edu/~elr/Robinson/XRbinary.pdf}}. The default phase zero of a light curve generated by XRBinary is inferior conjunction of the Roche-lobe filling secondary (i.e., the accretion disk around the white dwarf is occulted by the red component), which is in accord with the phases specified in eclipsing CVs. Compared with the typical CV eclipse light curve with a narrow and deep white dwarf eclipse \citep[e.g.][]{lit14,mca15}, all four light curves derived by \cite{dai16,dai17} only show wide and shallow modulations with an amplitude of several hundredths up to tenths of a magnitude. Note that the phases of the obtained light curves are arbitrary. The three DN (J0632+2536, RZ\,Leo and TW\,Vir) light curves clearly show double-hump modulations with a nearly constant phase difference of $\sim$0.5 between the two minima dips. This means that the light minima at the lower (primary dip) and higher (secondary dip) flux levels are at phases 0.5 or zero, respectively. In principle, an irradiation effect is significant in CV systems and commonly results in a higher flux level at the phase 0.5 dip than at phase zero. But, phase zero of the double-hump modulation cannot be simply identified from this flux difference due to possible changes caused by hotspots on the disk. Therefore, for each DN with a double-hump modulation discussed in this paper, two phased light curves corresponding to phase zero at the primary and secondary dips respectively, were analyzed to search for a final convergent solution. Since the secondary dip in TW\,Vir cannot be accurately measured due to large scatter in the data, the primary maximum was set to be at the reference phases 0.75 and 0.25, guaranteeing that minima would occur around phases 0.5 and zero. For the pre-CV WD\,1144+011, phase zero was set to be the minimum of the light curve since the light curve shows a single-hump modulation. 

The four K2 light curves are expressed in Simple Aperture Photometry (SAP) flux (i.e., electrons per second), while the light curves calculated by XRBinary are given in ergs per second. Hence, the K2 light curves were normalized before the calculations of XRBinary. Since the huge number of data points in the observed light curves (i.e., the number of data points in the K2 light curves observed in long cadence (LC; 30\,min sampling) and in short cadence (SC; 1\,min sampling) are more than 10$^{3}$ and 10$^{5}$, respectively.) can take a long calculation time with XRBinary, the K2 phased light curves are moderately binned with a uniform phase resolution 0.01. \cite{dai17} demonstrated that the orbital modulations of the three CVs are stable, and the large amplitude dispersion of $\sim$0.43\,mag in the phased light curve of RZ\,Leo \citep[Figure 5 of][]{dai16} is only caused by a uniform drift of the system light. Thus, while the phased light curve of TW\,Vir has the highest stability \citep{dai17}, all four binned and normalized K2 light curves represent their orbital modulations. 

\section{Synthesis Methods for the Light Curves}

Although the program XRBinary was initially designed for calculating the light curves of low-mass X-ray binary stars (LMXBs) \citep{gom15}, it can also be used to model low-inclination CVs based on the following two reasons. First, XRBinary sets the primary star as a much smaller sphere than all other dimensions in the binary system and is unresolved by any of the grids used in calculating the light curves. Like the neutron star in LMXBs, the white dwarf in both CVs and pre-CVs is very small compared with the companion red dwarf. Second, XRBinary simply assumes that the primary star only emits black body radiation. This is a reasonable assumption for low-inclination CVs and pre-CVs since the flux contribution from the visible white dwarf is approximately constant. XRBinary may not be appropriate for accurately modelling complicated eclipse light curves, but should be able to reproduce low-inclination CV light curves. In fact, XRBinary is a powerful tool to analyze the accretion disk around the central compact star by constructing a complex accretion disk model consisting of a disk, a disk rim, a disk torus and an inner disk. \cite{rat13} have successfully applied XRBinary to reproduce an ellipsoidal light curve of the CV CXOGBS J174444.7-260330 in a low state.

In this paper, a standard CV model (i.e., a semi-detached close binary system with a Roche-lobe filling red dwarf and, in the case of the 3 DN, an accretion disk around the white dwarf) with a set of geometric and physical parameters is applied to model the phased light curves of four low-inclination systems. Based on a complete set of preset parameters, XRBinary calculates a theoretical light curve and a relative measurement of goodness of fit, $\chi^{2}$ (i.e., the variance between the calculated and observed light curves). Taking advantage of the improved Nelder-Mead method (i.e., Multi-directional Search method \citep[MDS][]{nel65,tor89,gen08}), which was successfully applied to fit the O-Cs of AM Her \citep{dai13} and UZ\,For \citep{dai10b}, we developed a new program called NMfit to carry out a search for the best binary model within a given parameter space. After obtaining the best binary model, NMfit sets a series of small deviations around each of the parameters to test the variations in $\chi^{2}$ caused by these deviations. Note that when an adjustable parameter is intentionally set to deviate from its optimal value, the other parameters are fixed. Moreover, this optimisation method ignores degeneracies between parameters, which is a big issue. Since the best binary model means the minimal $\chi^{2}$, the deviations give rise to an increase of $\chi^{2}$. When $\chi^{2}$ increases to 5\% larger than the minimal value, this tested parameter deviation is arbitrarily regarded as the uncertainty. All adjustable parameters are tested one-by-one for estimating their errors. The error estimates based on $\chi^{2}$ are only lower limits to the true uncertainties in the derived parameters, and the true uncertainties could be quite large, since the physics of compact binaries is much more complicated and XRBinary arbitrarily makes many tacit assumptions (e.g. the disk is modelled as an object with sharp, well-defined edges and surfaces, and the disk surface elements emit and absorb like black bodies). The uniform design proposed by \cite{fan80} is used to decide the initial parameter set before the search iterates. A good initial parameter set is crucial for the success of the iterations. In addition, since Phoebe \footnote{The version of Phoebe used for the CV plotting is 2.0a2.} is able to calculate and plot an accretion disk around a compact star \citep{prs16}, it was used to visualize the system configuration based on the best-fitting parameters derived by NMfit and XRBinary. 

\section{Model Parameters Preset in XRBinary}

For generating light curves, there are a total of 23 model parameters listed in Table 1, which are divided up into 7 fixed parameters and 16 adjustable parameters. Our assumptions for these parameters are as follows.

1. Although the input and output light curves of XRBinary are normalized, the typical luminosity of the white dwarf can provide an important reference for the system luminosity of the output CV model. Since the mass range for isolated white dwarfs is 0.3\,M$_{\odot}$ to 1.4\,M$_{\odot}$ and over 30\% of white dwarfs are centered on $\sim$\,0.56\,M$_{\odot}$ \citep{ber92,pro98}, the white dwarf masses of all four objects were searched in this mass range. By using the relation L$_{wd}$=4$\pi$R$_{wd}^{2}\,\sigma$\,T$_{wd}^{4}$, the white dwarf radius R$_{wd}$ is solely dependent on the white dwarf luminosity L$_{wd}$ and temperature T$_{wd}$. According to the white dwarf mass-radius relation shown in Fig. 1 \citep[e.g.][]{woo90,pro98}, L$_{wd}$ can only be determined by T$_{wd}$ for a given M$_{wd}$. During the iterations of NMfit, T$_{wd}$ calculated from the adjustable parameter L$_{wd}$ is preset to be a fixed parameter.

2. In order to improve the reliability of the model, a simple disk with minimal free parameters consisting of up to two hotspots (one at the vertical side of the edge of the disk (es) and the other one on the surface of the disk (ss)) is used to describe the low-inclination CV light curves. This CV quiescent disk was assumed to normally extend to the surface of the white dwarf (i.e., R$_{in}$\,=\,R$_{wd}$). A boundary layer was not considered since it dominates the flux at ultraviolet wavelengths rather than at optical wavelengths. A hotspot on the vertical side of the edge of the disk can be completely described by using 3 parameters: a uniform temperature T$_{es}$, the centering phase $\zeta_{esmid}$ and the full width $\zeta_{eswidth}$. Considering that the other hotspot on the disk surface can be visible and modulate the light curves of low-inclination CVs, this hotspot is described by using five parameters: $\zeta_{ssmin}$, $\zeta_{ssmax}$, R$_{ssmin}$, R$_{ssmax}$ and T$_{ratio}$. The first four parameters are the boundaries of the hotspot, which are the angles and radii over which the hotspot extends, respectively. The last parameter T$_{ratio}$ is a fractional change in T$_{disk}$ (i.e., T$_{ratio}$\,=\,T$_{ss}$/T$_{disk}$).

3. A different temperature distribution may exist in different quiescent accretion disks. The temperature distribution is assumed to be a power law in disk radius R$_{disk}$, i.e., T$_{disk}\,\propto\,R_{disk}^{\xi}$. For a steady-state disk, typically $\xi$\,=\,-0.75 \citep{wad98}. However, subsequent papers suggested that accretion disks in CVs have much flatter temperature profiles with $\xi$\,$>$\,-0.75 \citep[e.g.][]{mar99,oro03}. According to the disk instability model \citep[DIM,][]{osa13a,osa13b,osa14}, the quiescent DN disk is an optically thin and cool disk with a flat radial temperature profile. Observations of various quiescent DN (e.g., Z Cha \citep{woo86}, OY Car \citep{woo89}, V2051 Oph \citep{rut16}) confirm that their disk temperature distributions are much flatter than the prediction of T$_{disk}\,\propto\,R_{disk}^{-0.75}$. We initially assume $\xi$\,=\,-0.15, but allow this value to be adjusted by NMfit.

4. The adjustable parameter L$_{d0}$ indicating the disk luminosity is only calculated from the temperature distribution of the disk. This parameter should satisfy a default condition that the temperature at the inner edge of disk (i.e., T$_{in}$) cannot be much larger than 10$^{4}$\,K, since the average temperature of observed quiescent disks is below 10$^{4}$\,K, which is consistent with the typical temperature range of an accretion disk (from several 10$^{3}$\,K to 10$^{4}$\,K) predicted by the limit-cycle oscillation between hot and cold states \citep{las01}.

5. Since both theories and observations suggest that the accretion disk around a white dwarf is geometrically thin \citep[e.g.][]{pri81,fra92,wad98,mar99}, the height of the disk, H$_{disk}$, is assumed to obey a power low with a power index H$_{pow}$=1.1 for all four objects, but H$_{edge}$ is set to be adjustable.

6. Since the components of the model calculated in XRBinary (e.g., two component stars, the disk and hotspots on the disk) are independent, we can freely construct the necessary components to model the observed light curves. Three types of models, as listed in Table 2, were attempted to verify the necessity of a disk around the white dwarf and the hotspots on the disk. Model-0 is a detached binary model without a disk. Model-1 contains an accretion disk with a hotspot at the edge of the disk. Model-2 adds a second hotspot on the disk surface. The relative flux contributions in percentage from the different model components to the synthetic light curves calculated by XRBinary are shown in Fig. 2. Their zero points are listed in Table 3.

\section{Irradiation effect calculated in NMfit}

The irradiation effect is only calculated by XRBinary when setting the parameter ``IRRADIATION" to be ``ON". Since a luminous secondary may weaken the illumination from the vicinity of the white dwarf and the disk, the parameter ``IRRADIATION" is preset to be ``OFF" if the summation of the luminosity of the white dwarf and the disk (i.e., L$_{wd}$+L$_{d}$) is smaller than the luminosity of the secondary L$_{rd}$. Although XRBinary can calculate the heating effect due to the irradiation, the output parameter T$_{rd}$ of XRBinary is simply derived from L$_{rd}$ by using the formula $L_{rd}\,\sim\,4\pi R_{rd}^{2}\,\sigma\,T_{rd}^{4}$. This means that T$_{rd}$ represents the temperature of the secondary averaged over its entire surface, which can be significantly higher than the true night side temperature of the secondary if irradiation in the systems is significant. Here, the night side is the surface of the secondary facing away from the WD and disk, and the day side is the surface facing the WD and disk.

\cite{dav92} found that the size of the irradiation region is usually as large as the entire day side surface of the secondary star. The numerical work by \cite{kir82} suggested that the difference in the effective temperature between the day and night sides of the secondary is very large. Considering that the typical difference may reach close to 10$^{4}$\,K \citep{war03}, irradiation would be easily observed if this tremendous disparity of the effective temperature exists. Therefore, an estimation of the size of the irradiation region on the surface of the secondary can help to understand the flux contribution from the irradiation region in both CV and pre-CV systems. The effective temperature of the secondary at the night side can be specified by the semi-empirical CV donor sequence \citep{kni06,kni11}. The derived T$_{rd}$ can be regarded as a lower limit to the effective temperature of the irradiation region. 

Due to orbital rotation, our viewing angle of the irradiated region can change with the orbital phase. Assuming that the irradiation region on the leading side of the star \footnote{The secondary is assumed to be spherical to simplify the calculations of the irradiation effect.} is axisymmetric with the line between the center points of two component stars, the flux ratio should reach a maximum between phases 0.5 and 0.0, i.e., F$_{irr}\,=\,f_{0.5}/f_{0.0}$, where f$_{0.5}$ and f$_{0.0}$ are the irradiation flux at phases 0.5 and 0.0, respectively. The details of the calculations of F$_{irr}$ are described in Appendix A. By investigating the variations in F$_{irr}$ along with a normalized area of the irradiation region (i.e., S$_{irr}\,=\,A_{irr}/A_{star}$, where A$_{irr}$ and A$_{star}$ are the areas of the irradiation region and the whole star, respectively), a relation between F$_{irr}$ and S$_{irr}$ based on the phycial model derived by XRBinary and NMfit can be estimated and used to compare with the observed flux ratio F$_{obs}$ between phases 0.5 and 0.0.

\section{Comparisons with Gaia results}

Since the K2 light curves of the four binary systems are from the broad bandpass listed in Table 1, their K2 magnitudes (hereafter Kp$_{2}$) translated by \cite{dai16} are commonly different from their apparent visual magnitudes. Assuming this magnitude difference to be a systematic error, the calculated magnitudes of the four systems (hereafter CKp$_{2}$) based on the luminosities of the models can be set to be a reference parameter for the comparisons with the results from the Gaia mission \citep{gai16}. Table 4 lists the distances of all four systems based on the absolute stellar parallax in the Gaia database \citep{gai18} \footnote{\url{https://gea.esac.esa.int/archive/documentation/GDR2/Miscellaneous/sec_credit_and_citation_instructions/}}. CKp$_{2}$ can be estimated from the Gaia distance D$_{g}$ derived from the Gaia parallax \citep{lur18}, and the system luminosity L$_{all}$, which is summed for all model components and corresponds to the zero point of the normalized K2 light curves, by using the following formula,
\begin{equation}
CKp_{2}\,=\,2.5\,\log\left[\frac{L_{bol0}}{L_{all}}\right]\,-\,BC_{v}\,+\,5\,\log\left[\frac{D_{g}}{10\,pc}\right],
\end{equation}
where L$_{bol0}\,=\,3.0128\times10^{35}$\,erg/s \footnote{A zero point luminosity corresponds to an absolute bolometric magnitude scale (i.e., M$_{bol}$=0) recommended by the IAU 2015 Resolution B2.}, and BC$_{v}$ is a bolometric correction of main sequence star in the typical Johnson-Cousins V band with the solar log\,g, [Fe/H] and [$\alpha$/Fe]. Note that BC$_{v}$ corresponding to the derived T$_{rd}$ can cause an uncertainty of CKp$_{2}$ for CV and pre-CV systems, since BC$_{v}$ is improperly used for the white dwarf and disk. This calculated uncertainty may explain the discrepancy between CKp$_{2}$ and Kp$_{2}$.

Considering that BC$_{v}$ is a model-dependent quantity with many observational constraints, several numerous tabulations are provided in the literature. All three popular BC$_{v}$ tables respectively proposed by \cite{flo96}, \cite{bes98} and \cite{cas14} were used to calculate CKp$_{2}$ to minimize our errors. The relations of BC$_{v}$ and T$_{eff}$ plotted in Fig. 3 clearly show that the three BC$_{v}$ tabulations significantly differ for cool stars with T$_{eff}\leq$\,4,000\,K. Inspection of Fig. 3 indicates that BC$_{v}$ of RZ\,Leo, TW\,Vir and WD\,1144+011 are obviously different for each table. Consequently, the difference between the three BC$_{v}$ tabulations for the same T$_{eff}$ is regarded as the uncertainty of BC$_{v}$. Compared with the two early BC$_{v}$ tables of \cite{flo96} and \cite{bes98}, the updated BC$_{v}$ given by \cite{cas14} is moderate for cool stars. Therefore, they were used to calculate CKp$_{2}$ for all four systems to compare with the K2 magnitudes at the zero point of the normalized K2 light curves \citep{dai16}. The interpolated BC$_{v}$ and the errors used for calculating CKp$_{2}$ for all four systems are listed in Table 4.

\section{Modelling Results and Discussions}

\subsection{J0632+2536}

This is a poorly studied DN with several DN outbursts in 2009 and 2012 \citep{kor12,mas12,ohs12}. \cite{dai16} showed its double-hump light curve extracted from the K2 data archive and several quiescent spectra with strong double peaked Balmer emission lines obtained from the Large Binocular Telescope (LBT). Based on these spectra and the phased K2 light curve, the secondary of J0632+2536 is thought to be a K5V star and the orbital inclination cannot be lower than 50$^{\circ}$. According to the MK spectral classes \citep{cox00}, the initial temperature and mass of the secondary are set to be 4,410\,K and 0.67\,M$_{\odot}$, respectively. Several high-precision measurements of white dwarf masses in CVs \citep{kni06,kni11,zor11} indicate that the mean CV white dwarf mass is usually larger than that of isolated white dwarfs \citep{ber92,kep07}. Thus, for J0632+2536, which lacks an accurate measurement of the white dwarf mass, a mean CV white dwarf mass of 0.83\,M$_{\odot}$ was preset to be the initial parameter (i.e., M$_{wd}$=0.83\,M$_{\odot}$). Finally, \cite{urb06} used IUE spectra of 53 quiescent DN to show that the temperature of a white dwarf, T$_{wd}$, in a CV with an orbital period above the period gap is approximately 25,793\,K \citep{sio99,urb06}. We assume this value for the temperature of the white dwarf in J0632+2536 since its orbital period of 0.314478\,day \citep{dai16} is above the gap.

A large number of trials calculated using XRBinary and NMfit suggest that a convergent solution using a normal CV configuration cannot be achieved when setting phase zero to the primary dip of the light curve. Hence, phase zero was set to be the secondary dip. Model-1 was used to model the phased light curve of J0632+2536. At first, four parameters (M$_{wd}$, T$_{wd}$, mass ratio q$_{orb}$ and T$_{rd}$) were fixed, and the other nine parameters consisting of the orbital inclination i$_{orb}$ and eight parameters of the accretion disk were set to be adjustable. By using NMfit, a preliminary accretion disk model indicated that the inner radius of the disk is basically close to the white dwarf radius, so we then fixed the parameters $\xi$ and R$_{in}$ for deriving the uncertainties of the other 11 parameters. Since $\xi$ is an insensitive parameter, its uncertainty cannot be obtained. By using Phoebe 2.0, a 2D CV configuration at phase 0.75 shown in Fig. 2 visually indicates that J0632+2536 has a large and thick accretion disk with a small hotspot at the edge of the disk. Since the orbital inclination of J0632+2536 is not high enough for the white dwarf to be occulted by the secondary, the flux contribution from the white dwarf is constant and does not vary with the orbital period, so this constant flux contribution is simply added to the ellipsoidal modulations caused by the secondary. Compared with the small white dwarf, the top panel of Fig. 2 clearly shows that the large and thick accretion disk can be partially eclipsed around phase zero. Moreover, the hotspot at the edge of the disk, which has a temperature of 6,200 ($\pm$100) K and has a phase width of 0.033(2) at phase 0.844(2), only contributes a maximum of 5\% of the flux from the whole system. Since CKp$_{2}$ of J0632+2536 listed in Table 4 is almost equal to the corresponding Kp$_{2}$, the obtained physical model of J0632+2536 is compatible with the Gaia distance. However, the T$_{eff}$ of J0632+2536 shown in the Gaia database is obviously higher than that derived from the LBT spectra \citep{dai16} and the derived T$_{rd}$ listed in Table 5.

The region where the mass transfer stream intersects the accretion disk (i.e., the hotspot at the edge of the disk) is simply assumed to be a rectangle with a uniform temperature T$_{es}$ by XRBinary. A mass transfer rate $\dot{M}_{rd}$ (hereafter, the parameters with the subscript rd related to the secondary star) can be roughly estimated by using the following formula,
\begin{equation}
\dot{M}_{rd}\,\simeq\,\frac{L_{acc}\,R_{out}}{G\,M_{wd}},
\end{equation}
where L$_{acc}$ is the luminosity of the hotspot at the edge of the disk. The mass transfer rate can be estimated to be $\dot{M}_{rd}\sim2.5(\pm0.2)\times10^{-10}\,M_{\odot}$/year, corresponding to a mass loss timescale (i.e., $\tau_{\dot{M}}\,\sim\,M_{rd}/\dot{M}_{rd}$) of 2.8$\times10^{9}\,$year. The top right panel of Fig. 4 indicates that M$_{rd}$\,=\,0.7(1)\,M$_{\odot}$ is closer to the semi-empirical mass-period relation of \cite{war03} than that of \cite{smi98}. Furthermore, the secondary radius R$_{rd}$\,=\,0.81 is consistent with the radius-period relations shown in the bottom right panel of Fig. 4. The thermal (or Kelvin-Helmholtz) timescale of the secondary (i.e., $\tau_{kh}\,\sim\,GM_{rd}^{2}/(L_{rd}R_{rd})$) can be calculated to be 7.6$\times10^{7}$\,year, as listed in Table 6, is smaller than the derived $\tau_{\dot{M}}$. This means that the mass transfer via the L1 point is slow, and the secondary is always able to maintain thermal equilibrium. Therefore, the secondary of J0632+2536 is almost indistinguishable from an isolated main sequence star of the same mass. The derived mass and radius shown in the bottom left panel of Fig. 4 confirm that the secondary of J0632+2536 is a typical K5V star \citep{dai16}. Inspection of the top panel of Fig. 2 indicates that the ellipsoidal modulations of this K5 star dominate the observed double-hump modulations. Although the top left panel of Fig. 4 indicates that T$_{rd}$\,=\,4,540(80)\,K is nearly consistent with a normal K5V star based on the MK spectral classes \citep{cox00}, this derived T$_{rd}$ is around 350\,K higher than the prediction of the semi-empirical CV donor sequence \citep{kni06,kni11}. The most likely explanation for this discrepancy may be due to irradiation of the K5V star. Since the irradiation effect is not calculated by XRBinary as L$_{rd}>L_{wd}+L_{d}$, we investigated its strength in more detail. Based on our best-fitting model, the maximal F$_{irr}$ is around 1.08 as shown in Fig. 5. As long as 20\% of the irradiation region can be occulted by the disk at phase 0.5, the observed flux ratio of J0632+2536 (i.e., F$_{obs}\sim$0.87) can be explained. Due to our derived large disk and moderate orbital inclination, a partial eclipse of the irradiation region at phase 0.5 can be expected. 

\subsection{RZ\,Leo}

\cite{dai16} confirmed that RZ\,Leo is a short-period DN with an orbital period of 0.07603\,day similar to that derived by \cite{pat03} and \cite{kat09}. \cite{ish01} and \cite{men99} indicated that the secondary of RZ\,Leo should be a normal red dwarf, possibly a main sequence M0 star. Although this spectral type cannot be supported by the mass ratio of 0.14 derived from its superhump excess of 0.033 \citep{ish01}, we speculated that RZ\,Leo should consist of a massive white dwarf and a normal red dwarf with a small mass of $<$\,0.2\,M$_{\odot}$. Thus, we assumed that the initial M$_{rd}$ used in NMfit is 0.15\,M$_{\odot}$, which means M$_{wd}$\,=\,1.1\,M$_{\odot}$ and T$_{rd}$\,$\sim$\,3,500\,K. \cite{szk17} and \cite{dai16} detected a short white dwarf spin period of $\sim$\,220\,s classifying RZ\,Leo as a member of the IP subtype of CVs. From an Ultraviolet (UV) study, \cite{pal17} determined that the white dwarf had a temperature of 15,014($\pm$638)\,K and contributed 83\% of the UV flux. Since the humps of the model-1 light curve are located exactly at phases 0.25/0.75, model-2 with an extra hotspot on the surface of disk can be used to describe the offset secondary hump of RZ\,Leo. For the phased light curve with phase zero at the primary minimum, a convergent solution was found. However, inspection of the bottom left panel of Fig. 4 shows that the derived secondary mass and radius obviously deviate from the predicted mass-radius relation of the semi-empirical CV donor sequence \citep{kni06,kni11}. Based on the MK spectral classes \citep{cox00}, the derived radius is significantly smaller than that of a normal isolated main sequence star. Furthermore, the mass and radius also conflict with the semi-empirical mass-period and radius-period relations of CV secondary \citep{war03,smi98} shown in the right two panels of Fig. 4.

Instead, we created a model with phase zero at the secondary dip of the light curve. Initially, we fixed the inner accretion disk radius to the radius of the white dwarf. However, a convergent solution could not be found. Instead, an accretion disk model with two hotspots and R$_{in}$ much larger than R$_{wd}$ was found to fit the data reasonable well. R$_{in}>R_{wd}$ can be interpreted as a truncation of the accretion disk far from the WD, consistent with the IP classification of RZ Leo by \cite{szk17} and \cite{dai16} This means that the physical structure of RZ\,Leo may be more complex than the simple disk model calculated by XRBinary, as accretion curtains near the magnetic poles of the white dwarf may be involved. This may be the reason that the best-fitting light curve derived by XRBinary cannot perfectly describe the secondary hump of RZ\,Leo. Based on this disk model, CKp$_{2}$ calculated by using Equation 1 is around 1.5 magnitudes larger than Kp$_{2}$. This discrepancy is likely caused by the lack of an accretion curtain in the disk model. In the future, a complete IP model including the light from an accretion curtain should help correct this difference. In spite of the deviation in the secondary hump and CKp$_{2}$, limits to two key parameters of the disk (R$_{in}$ and $\xi$) can be obtained from the double-humped K2 phased light curve.

According to the typical disk-field interaction models \citep[e.g.][]{pri72,gho78,whi88}, the corotation radius of a magnetic white dwarf, R$_{co}$, can be calculated using,
\begin{equation}
R_{co}\,=\,f_{co}M_{wd}^{1/3}P_{rot}^{2/3},
\end{equation}
where $f_{co}=(GM_{\odot}/4\pi^{2}R_{\odot}^3)^{1/3}$, R$_{co}$ and M$_{wd}$ are in solar units, P$_{rot}$ is the spin period of the magnetic white dwarf in unit of seconds. By using the parameters listed in Table 5 and the spin period of 220\,s \citep{szk17,dai16}, R$_{co}$ of RZ\,Leo can be estimated to be 0.085\,R$_{\odot}$, which is 17 times larger than R$_{wd}$\,=\,0.005\,R$_{\odot}$. Considering that the accretion process in RZ\,Leo is steady \citep{dai17}, R$_{in}$ is required to be almost equal to R$_{co}$. However, R$_{in}$\,=\,0.211(2)\,R$_{\odot}$ is much larger than R$_{co}$. In a typical IP \citep{gho78}, R$_{co}$ is regarded as the inner radius of an unperturbed disk, and the transition region of a truncated disk (i.e., located between R$_{co}$ and R$_{in}$) may be actually invisible in the optical band due to a possible disruption of the accretion flow caused by the magnetosphere of the magnetic white dwarf. Since a hotspot on the disk surface always shows R$_{ssmin}<R_{in}$ during the iterations of NMfit, the parameter of R$_{ssmin}$ is fixed to be equal to R$_{in}$. The high effective temperature of this hotspot indicated by T$_{ratio}$\,=\,7.2 indicates that more than 97\% of the luminosity of the disk is from this slim and extended hotspot. Moreover, this hotspot is regarded to be a plausible second impact region of an inward and overflowing stream striking onto the magnetosphere of the magnetic white dwarf. This region was previously reported in several IP systems (e.g., EX\,Hya in outburst \citep{hel89} and QZ\,Vir in quiescence \citep{sha84}. Note that all three systems are unusual short period (under the gap) IPs.). Compared to this hotspot, the other hotspot at the edge of the disk is much smaller and cooler. Both edge and surface hotspots are located at the phases 0.723(1) and 0.47(2), respectively. This disk configuration derived in quiescence is consistent with the disk of EX\,Hya during outburst, with two hotspots at similar phases \citep{hel89}. This disk configuration may be common for IP systems. A low-luminosity truncated disk with L$_{d}=2.3\times10^{30}\,erg\,s^{-1}$ (i.e., the luminosity contribution of the disk is less than 10\% of the whole CV system) may be a straightforward conclusion for an IP system like RZ\,Leo due to the lack of a hot inner part of the disk. Compared with the other four sources of light in RZ\,Leo (i.e., the two component stars and two hotspots on the disk), the relative flux contributions from this truncated disk shown in Fig. 2 are almost negligible (close to zero).

The best-fitting CV model indicates that RZ\,Leo has a moderate orbital inclination of 61.0$^{\circ}$($\pm$0.9$^{\circ}$) and contains a massive white dwarf and a low mass red dwarf. The mass ratio, q$_{orb}=0.080(6)$ is within the error bar of that estimated from its superhump excess \citep{ish01}. The secondary mass and radius (i.e., M$_{rd}$\,=\,0.101(8)\,M$_{\odot}$ and R$_{rd}$\,=\,0.16\,R$_{odot}$) are not only in accord with an isolated red dwarf later than M5V, but also support the semi-empirical mass-period and radius-period relations of CV secondaries \citep{war03,smi98}. The small mass transfer rate, $\dot{M}_{rd}\simeq\,3.01(\pm0.07)\times10^{-12}\,M_{\odot}$/year calculated by using Equation 2, seems to explain the small and cool hotspot at the edge of the disk shown in the disk model. Recently, \cite{dub18} showed the average mass transfer rates of a sample of $\sim$130 CVs with a parallax distance in the Gaia DR2 catalogue. The estimated mass transfer rate of RZ\,Leo lies well within the region where a stable, cool disk exists within the system. This explains a lack of DN outbursts for RZ\,Leo. $\tau_{\dot{M}}\,\sim\,3.4\times10^{10}$\,year calculated from the mass transfer rate is larger than $\tau_{kh}$ of the secondary. The derived secondary mass and radius further confirm that the secondary of RZ\,Leo is in thermal equilibrium, since they are consistent with the mass-radius relation of the isolated main sequence stars shown in the bottom left panel of Fig. 4. Since T$_{rd}$\,=\,3,710($\pm$30)\,K suggests a slightly earlier spectral type than M5V, this higher T$_{rd}$ may be attributed to irradiation of the companion. Due to L$_{rd}>L_{wd}+L_{d}$, the irradiation effect is not calculated by XRBinary. However, the maximal F$_{irr}\sim$1.54 shown in Fig. 5 indicates a significant irradiation effect in RZ\,Leo. A prominent irradiation effect can be expected in an IP system since the white dwarf primary is not embedded in the truncated thin disk. Furthermore, the small disk of RZ\,Leo implies that the irradiation region is visible at phase 0.5 despite the similar orbital inclination to that of J0632+2536. Although F$_{obs}\sim$\,0.93 does not seem to support a large irradiation effect, the presence of an accretion curtain and a brighter second impact region complicate an accurate calculation of the irradiation effect in RZ\,Leo.

\subsection{TW\,Vir}

By using infrared photometry, the secondary is estimated to be a M3V star with a mass of 0.43\,M$_{\odot}$ and a radius of 0.48\,R$_{\odot}$ \citep{mat85}. According to the MK spectral classes \citep{cox00}, the initial temperature of the secondary was set to be 3,367\,K. Based on the mass ratio of 0.44 \citep{sha83}, the white dwarf mass of TW\,Vir is calculated to be 0.98\,M$_{\odot}$. Since there is not any accurate white dwarf temperature obtained from UV spectra \citep{cor82,szk85,ham07}, and TW\,Vir is a long period (above the gap) system with an orbital period of 0.182682(3)\,day derived from the K2 data \cite{dai17}, we set the same initial white dwarf temperature as that of J0632+2536. As the secondary hump of TW\,Vir is not located at orbital phase 0.25 or 0.75, model-2 is used for the calculations. Two phased light curves corresponding to phase zero at the primary and at the secondary dips were used to derive two disk models, respectively. In these two disk models, three parameters involving the positions of two hotspots on the disk in $\zeta$ direction (i.e., $\zeta_{esmid}$, $\zeta_{ssmin}$ and $\zeta_{ssmax}$) show nearly constant phase differences of 0.5, while all the remaining parameters are almost the same. Therefore, both models are not significantly different and the double-hump modulation of TW\,Vir is mainly caused by the two hotspots on the disk.

Since the $\chi^{2}$ of the model based on the phased light curve with phase zero around the secondary dip is slightly smaller than that with phase zero at the primary dip, the primary hump was set to be phase at 0.75 and the corresponding model is analyzed in the following discussion. The two best-fitting orbital parameters: q$_{orb}$=0.41($\pm$0.05) and i$_{orb}$=44.3$^{\circ}(\pm0.5^{\circ}$) listed in Table 5 are consistent with the previous results derived from the infrared and UV data \citep{cor82,mat85}. The derived mass ratio confirms the previous classification of TW\,Vir as a U\,Gem type DN \citep{oco32}, but the derived orbital inclination is smaller than that obtained by \cite{ham07}. The calculated CKp$_{2}$ listed in Table 4 suggests that the best-fitting model of TW\,Vir is consistent with the Gaia distance. The relative flux contributions from the hotspot at the edge of the disk around phase 0.75 shown in Fig. 2 almost perfectly reproduces the significant primary hump which lasts for over half of an orbit. Although the other hotspot appearing on the outer radius of the disk covers a long phase width of 0.42, the small T$_{ratio}^{ss}$=1.20($\pm$0.02) may explain its small relative flux contributions shown in Fig. 2. Like RZ\,Leo, the relative flux contribution from the disk is close to zero. The derived mass and radius of the secondary fits with the semi-empirical secondary mass-period and radius-period relations \citep{war03,smi98}, indicating that the spectral type of the secondary is close to M2V, which is similar to the result from infrared photometry \citep{mat85}. However, compared with the K5V star of J0632+2536 and the M5V star of RZ\,Leo, the M2V star of TW\,Vir never shows large-amplitude ellipsoidal modulations. Instead, it gives rise to a small amplitude second hump in the phased light curve. By using Equation 2, the estimated $\dot{M}_{rd}$ listed in Table 6 indicates that $\tau_{\dot{M}}\sim9\,\tau_{kh}$, which is common for most CV secondaries \citep{pat84}. A somewhat oversized secondary of TW\,Vir relative to an isolated main sequence star of the same mass is not obtained, implying that $\dot{M}_{rd}$ may be overestimated.

Like J0632+2536, T$_{eff}$\,=\,4850\,K listed in Table 4 is higher than the derived T$_{rd}$\,=\,4,000($\pm$40)\,K. This may imply that the higher T$_{eff}$ derived by Gaia is common for DN due to the possible contributions from hotter components in these systems (e.g., the white dwarf and accretion disk). Since T$_{rd}$ is obviously higher than the effective temperature of a M2V star shown in Fig. 4, we investigated the irradiation effect of the secondary in this system. Note that TW\,Vir is the only object with L$_{rd}<L_{wd}+L_{d}$. Accordingly, the calculation of irradiation was included in XRBinary. However, both light curves calculated with and without irradiation by XRBinary are almost identical. Hence, irradiation in TW\,Vir may be weak, which is also demonstrated by F$_{obs}\sim1$ with two minima in the light curve at almost the same flux level. Since L$_{d}$ calculated from a near-flat temperature distribution with a power index of -0.11($\pm$0.04) is nearly two times L$_{rd}$, and the total contributions of two hotspots are only 16\% of L$_{d}$, the weak irradiation effect of the secondary is overpowered in the light curve. Large variations in the disk luminosity may further weaken the ability to detect irradiation. There are two issues which further complicate our conclusions. The first is that the orbital modulation of TW\,Vir shown in Fig. 2 is only extracted from the part of the quiescent light curve around its superoutburst. The second is that the double-hump modulation does not always maintain stability and the secondary hump has a small amplitude and shows a large scatter in the unbinned light curve \citep{dai17}. Further analysis of its light curves at different times may reveal more details of the irradiation effect in TW\,Vir.

\subsection{WD\,1144+011}

This poorly understood variable star is classified as a DA+dMe binary from a single optical spectrum obtained by \cite{ber92}. For comparison with this spectrum obtained over twenty years ago, a new spectrum was taken on 2017 January 22 by using the BFOSC spectrograph attached to the Xinglong Observatory 2.16\,m telescope (XL 216, \cite{fan16}). This is a better optical spectrum showing some changes from the original one taken in long ago. The details concerning our spectrum are presented in Appendix B. A synthetic analysis based on the single-hump light curve extracted by \cite{dai16} can test the possibility of the existence of a disk around the primary white dwarf of WD\,1144+011. Considering that WD\,1144+011 is a long orbital period system with an orbital period of 9.81\,hr \citep{dai16}, the initial parameters of the white dwarf are set to be the same as those of J0632+2536. Combined with the parameters of a M dwarf, all three models were used to attempt to reproduce the stable single-hump modulation that is apparent in the K2 light curve.

\subsubsection{Model without a disk}

In principle, model-0 can only produce ellipsoidal modulation caused by the orbital motion of the Roche-lobe filling secondary. The asymmetry shown in the single-hump modulation of WD\,1144+011 cannot be explained by a pure ellipsoidal modulation derived from model-0. Moreover, the irradiation effect in WD\,1144+011 calculated by XRBinary cannot resolve this problem, since its long orbital period may imply a bright secondary like J0632+2536 (i.e., L$_{rd}>L_{wd}$). In spite of this, an irradiation region on the surface of the secondary was added to allow for an asymmetrical single-hump modulation (i.e., a bump on the rise to the maximum). We assumed this irradiation region to be a circular bright starspot on the secondary, which can be easily calculated by XRBinary. Since the single-hump modulation of WD\,1144+011 observed in K2-C1 is very stable (lasting at least a complete campaign period of $\sim$\,3\,months \citep{dai16}), a stable region due to irradiation of the secondary may be more plausible to explain this long-term steady modulation than a variable starspot on an active M type dwarf.

In XRBinary, a starspot on the secondary can be described by the four parameters (i.e., $\theta^{sp}$, $\phi^{sp}$, R$^{sp}$ and T$_{ratio}^{sp}$) \footnote{$\theta^{sp}$ and $\phi^{sp}$ are the coordinates of the spot center in the spherical polar coordinate system. $\phi^{sp}$\,=\,0 direction lies in the orbital plane and points in the direction of motion of the secondary in its orbit. $\phi^{sp}$ increases in a right-handed sense about the $\theta^{sp}$\,=\,0 direction. R$^{sp}$ is the angular radius of the spot as seen from the center of the secondary star. T$_{ratio}^{sp}$ is the ratio of the starspot temperature to the local effective temperature of the unspotted star.}. For WD\,1144+011, this ``starspot" is assumed to be a hot spot (i.e., T$_{ratio}^{sp}>$1). All parameters including four component parameters (M$_{wd}$, T$_{wd}$, q$_{orb}$ and T$_{rd}$) and four parameters of the starspot were set to be adjustable in NMfit. The derived M$_{wd}=1.11(8)\,M_{\odot}$ is almost the same as that of TW\,Vir shown in Fig. 1, while T$_{wd}=27,100\,K$ suggests a hotter and brighter white dwarf. Compared with the three DN, WD\,1144+011 is the only object with a slightly cooler M dwarf (T$_{rd}=3,500(\pm100)\,K$) than the predicted mass-temperature relation of the semi-empirical CV donor sequence \citep{kni06,kni11}. This is consistent with T$_{eff}$\,=\,3,657.5\,K shown in the Gaia database. Compared with the higher T$_{eff}$ of the two DN J0632+2536 and TW\,Vir, this compatible T$_{eff}$ can be explained by the lack of a disk around the white dwarf. Since the companion star in WD\,1144+011 is cooler than the secondaries in the other systems discussed here, it has the largest uncertainty in BC$_{v}$, as shown in Fig. 3. This is reflected in the large range of derived CKp$_{2}$ (14.8$\sim$16.2\,mag) which is consistent with the corresponding Kp$_{2}$ value listed in Table 4. Like TW\,Vir, the ellipsoidal modulations caused by the M star never dominates the orbital modulation. The mass and radius of the secondary deviate from all the mass-radius, mass-period and radius-period relations shown in Fig. 4. Note that both semi-empirical mass-period and radius-period relations are only available in the period range of 1.3\,$\sim$\,9\,hr \citep{war03,smi98}. WD\,1144+011 with an orbital period of 9.81\,hr is beyond this period range. The bottom left panel of Fig. 4 shows that the inflation of the secondary is about 38\%. Besides the known inefficient mechanisms of donor bloating (e.g., tidal deformation, rotational deformation and irradiation of the secondary) discussed by \cite{kni11}, the deviation from thermal equilibrium of the secondary caused by the donor mass loss can result in a large donor inflation up to 20\%\,$\sim$\,30\% \citep{pat05,kni06}. However, this mass loss mechanism is still not enough to explain the derived inflated secondary star.

Although both light curves calculated with and without irradiation by XRBinary are almost identical due to L$_{rd}>L_{wd}$, the final $\chi^{2}$ of model-0$^{irr}$ is slightly smaller than that of model-0. Thus, the best-fitting parameters and their uncertainties listed in Table 7 are calculated using model-0$^{irr}$. The reproduced light curves and the corresponding 2D pre-CV configuration are shown in the top left and right panels of Fig. 6, respectively. This detached binary configuration with a low inclination of 14.3$^{\circ}(\pm0.4^{\circ}$) supports the previous classification of WD\,1144+011 as a pre-CV system (i.e., a DA+dMe detached system). Our model also shows that the orbital variation seen in the light curve is dominated by the hot spot on the secondary's surface. Compared with the three DN, the pre-CV WD\,1144+011 is the only object with F$_{obs}$\,=\,1.03 larger than 1. Thus, T$_{ratio}^{sp}=1.097(\pm0.003)$ combined with F$_{obs}$ suggests that the calculated irradiation region may be a good representation of the physical picture of WD\,1144+011. The effective temperature of the ``starspot" can be estimated to be around 3,900\,K based on T$_{ratio}^{sp}$. A relation between F$_{irr}$ and S$_{irr}$ shown in Fig. 5 indicates weak irradiation in WD\,1144+011, similar to J0632+2536, and S$_{irr}\sim$\,0.5 corresponding to F$_{obs}$=1.03 is consistent with the size of the ``starspot" calculated from R$^{sp}=27.7^{\circ}(\pm0.7^{\circ})$. Due to the lack of an accretion disk, this irradiation region is much smaller than the typical size of an irradiation region in a CV \citep{dav92}.

\subsubsection{Model with a disk}

Since an extra light source is imperative to explain the asymmetrical single-hump modulation, the model-1 and model-2 were investigated. All adjustable parameters and their uncertainties are listed in Table 7. $\dot{M}_{rd}$ is estimated for both models, and is given in Table 6. The white dwarf mass of model-1 is smaller than the prediction of the average white dwarf mass in a CV with the same orbital period \citep{rit86,zor11}. T$_{wd}$ from both of the disk models is consistent with T$_{wd}$ from the non-disk model, suggesting that WD\,1144+011 does contain a hot white dwarf. The final $\chi^{2}$ of the two disk models are almost the same as that of the non-disk model-0. All three models imply an oversized secondary with generally consistent masses and radii. The top left panel of Fig. 4 shows that the secondary of model-1 has an extremely high effective temperature which is over twice the average effective temperature of a M dwarf. Although the secondary of model-2 agrees with a G9V star based on the MK spectral classes \citep{cox00}, a spectral type of G9V contradicts with our spectrum. Compared with the Gaia distance of WD\,1144+011 listed in Table 4, both L$_{rd}$ for the model-1 and model-2 are too large. Hence, the consistency of the effective temperature may be only coincidental. In order to recheck the deviations in T$_{rd}$ in model-1 and model-2, T$_{rd}$ was fixed to be 3,500\,K in line with the secondary temperature found using model-0. However, no convergent solution could be obtained. In spite of L$_{rd}>L_{wd}+L_{d}$, irradiation was still included in these models. Like model-0, the light curves calculated with and without irradiation for the two disk models only show small discrepancies around phases 0.5 and 0.0. Both calculated F$_{irr}$ are far larger than F$_{obs}$. This means that the assumed large irradiation of the secondary is not supported by the K2 data. Compared with the pre-CV configuration obtained from model-0, the two models with an accretion disk are not convincing. As such, our modelling based on the K2 phased light curve provides further evidence for the classification of WD\,1144+011 as a detached pre-CV system.

\section{Conclusions}

\subsection{Synthetic codes}

Based on the light curve synthesis code XRBinary derived by E.L. Robinson, NMfit was developed to analyze the light curves of the four low-inclination systems: J0632+2536, RZ\,Leo, TW\,Vir and WD\,1144+011. All parameters of the best-fitting models and their uncertainties are estimated by NMfit. Additionally, Phoebe 2.0 was used to visualize the configuration of each system using each systems best-fitting parameters. Since phase zero is hard to identify in any low-inclination CV system with a double-hump modulation due to the lack of a significant eclipse feature, we tested models which had phase zero located at either the primary or secondary minimum, and chose the model which matched the observed light curve best. Except for WD\,1144+011 which had a single maximum in its phased orbital light curve, the derived CV models of the other three DN indicate that phase zero should be placed at the secondary minimum.

\subsection{Physical Models}

For J0632+2536 and TW\,Vir, the best-fit disk models show that the primary hump is mostly due to the hotspot at the edge of the disk, a key indicator of mass transfer via the L1 point. Another hotspot on the disk surface can explain the phase difference between the two humps of the double-hump modulations. For WD\,1144+011, the bright ``starspot" representing irradiation of the secondary star is responsible for the modulation seen in the optical light curve.

The lack of an accretion curtain in the disk model of RZ\,Leo implies that this model may not be appropriate for a comparison with the measured distance by Gaia. The derived physical models of the other three binary systems are consistent with the results from Gaia DR2. The flat power law index of the disk found in all three DN models ($\xi\sim$\,-0.11) is similar to previous observations \citep[e.g.][]{woo86,woo89,rut16} and supports the theory that a quiescent CV disk deviates from the temperature distribution of a typical steady-state disk. A low-luminosity accretion disk model of RZ\,Leo derived from its K2 light curve further confirms that RZ\,Leo is an IP system with two hotspots on a truncated disk. One of the hotspots on the disk surface contributes a significant fraction of the disk luminosity ($>97\%$ of L$_{d}$), and is located at the inner edge rather than the outer edge of the disk. This may be evidence of an impact region between an inward and overflowing stream and the magnetosphere of the magnetic white dwarf. Compared with RZ\,Leo, the small hotspot of J0632+2536 and two hotspots of TW\,Vir covering large phase ranges are only small contributors to the disk luminosity. Our spectrum of WD\,1144+011 with a relatively flat continuum and H$_{\alpha}$ emission supports its previous classification as a DA+dMe system \citep{ber92}. We note that WD\,1144+011 shows different flux levels in the continuum and emission lines. The model light curve based on the asymmetrical single-hump modulation requires an extra light source (i.e., a weak irradiation region of the secondary rather than a large hotspot at the edge of the disk) to explain the modulation of WD\,1144+011.

\subsection{The Secondaries}

The estimated $\dot{M}_{rd}$ for all four objects are within a range of $10^{-9}\sim10^{-12}\,M_{\odot}$/year. Except for the pre-CV WD\,1144+011 which contains an oversized secondary, the other three DN have secondaries in thermal equilibrium with masses and radii conforming to the semi-empirical CV donor sequence \citep{kni06,kni11} and MK spectral classes \citep{cox00}. The derived effective temperatures of all three DN are significantly higher than predicted. Hence a DN system containing a substantially hotter secondary may be a common feature rather than a peculiarity. This can be attributed to irradiation of the secondary, since T$_{rd}$ calculated from L$_{rd}$ is an average parameter which can be increased by irradiation. Compared with T$_{eff}$ listed in the Gaia catalog, the lower T$_{rd}$ of the two DN J0632+2536 and TW\,Vir may due to contamination from a hot white dwarf and disk. This is further supported by the T$_{rd}$ of the pre-CV WD1144+011, which is almost consistent with the Gaia T$_{eff}$. It should also be noted that the Gaia temperatures are determined from three broad bandpasses \citep{and18} and the DR2 releases notes urge caution in using them \footnote{\url{https://gea.esac.esa.int/archive/documentation/GDR2/pdf/GaiaDR2_documentation_1.0.pdf}}.

Although the double-hump modulation of J0632+2536 can be explained by the partial occultation of the irradiation region on the surface of the secondary due to a large disk and a moderate orbital inclination, investigation of irradiation in the other two DN implies that the effect of irradiation in a CV system is complicated and blended with other modulations. The flux contribution from the secondary of TW\,Vir is the lowest (i.e., L$_{rd}<L_{wd}+L_{d}$) among all four objects. Weak irradiation may exist in the DN TW\,Vir and the pre-CV WD\,1144+011. The former can be further tested by additional light curves obtained when the double-hump variation is evident, while the latter can be further checked by taking a time series of spectra over the course of the complete orbital period of 9.81\,hr.
 
\acknowledgments

This work was partly supported by CAS Light of West China Program, the Chinese Natural Science Foundation (Nos. 11133007 and 11325315), and the Science Foundation of Yunnan Province (No. 2016FB007). PS acknowledges support from NSF grant AST-1514737. M.R.K is funded through a Newton International Fellowship provided by the Royal Society. We acknowledge the support of the staff of the Xinglong 2.16m telescope. This work was partially supported by the Open Project Program of the Key Laboratory of Optical Astronomy, National Astronomical Observatories, Chinese Academy of Sciences. We would like to thank the anonymous referee for their comments and suggestions in improving this paper.

\software{IRAF \citep{tod86,tod93}, XRBinary (v2.4), NMfit (v1.0), Phoebe \citep[v2.0;][]{prs16})}

\clearpage

\appendix
\section{A simple calculation for the irradiation effect of the secondary in a low-inclination CV system}

Assuming that the secondary is a sphere and the irradiation region with a higher effective temperature of T$_{irr}$ is axisymmetric with the line between the center points of the two component stars, the area of the complete irradiation region S$_{irr}$ can be calculated by the following formula,
\begin{equation}
A_{irr}\,=\,2\pi R_{rd}^{2}\,(1-\cos{\theta}),
\end{equation}
where R$_{rd}$ and $\theta$ are the radius of the secondary and the half opening angle of the irradiation region shown in Fig. A, respectively. Therefore, $\theta$ should be in a range of 0$\sim\pi/2$. Due to the projection effect (i.e., the orbital inclination less than 90$^{\circ}$), the orbital modulation of the secondary will cause the fraction of the irradiation region seen by an observer to vary over the orbital period. The total flux at an orbital phase $\phi$ can be described as
\begin{equation}
f_{\phi}\,=\,A_{irr}^{'}(\phi)\,\sigma\,T_{irr}^{4}+[2\pi R_{rd}^{2}\,-\,A_{irr}^{'}(\phi)]\,\sigma\,T_{star}^{4},
\end{equation}
where A$_{irr}^{'}(\phi)$ and T$_{star}$ are the visible area of the irradiation region at the phase $\phi$ and the effective temperature of the unirradiated part of  the secondary, respectively. Note that Equation A2 never considers the occultation of irradiation region by the white dwarf or the disk. In principle, A$_{irr}^{'}$(0.0) is the minimum, while A$_{irr}^{'}$(0.5) is the maximum. Inspections of Fig. A indicates that the former can be estimated by the formula,
\begin{equation}
A_{irr}^{'}(0.0)\,=\,
\left \{
\begin{array} {l}
2R_{rd}^{2}\int_{i}^{\theta}\,\theta^{'}\cdot\sin{\psi}\,d\psi\quad\quad\theta\ge i\\
0\qquad\qquad\qquad\qquad\qquad\theta<i\\
\end{array}
\right.
\end{equation}
where i and $\psi$ are the orbital inclination and the second coordinate of the spherical coordinate system, respectively. Moreover, $\theta^{'}$ satisfies the following relation,
\begin{equation}
\frac{R_{rd}\sin{\psi}\cos{\theta^{'}}}{R_{rd}\sin{i}}\,=\,\frac{R_{rd}\cos{\psi}}{R_{rd}\cos{i}},
\end{equation}
By combining Equations A3 and A4, $A_{irr}^{'}$(0.0) can be expressed as,
\begin{equation}
A_{irr}^{'}(0.0)\,=\,
\left \{
\begin{array} {l}
2R_{rd}^{2}C_{irr}\quad\quad\theta\ge i\\
0\qquad\quad\,\qquad\theta<i\\
\end{array}
\right.
\end{equation}
where $\displaystyle{C_{irr}=\int_{i}^{\theta}\arccos{(\frac{\tan{i}}{\tan{\psi}})}\cdot\sin{\psi}\,d\psi}$. Based on A$_{irr}^{'}$(0.0), A$_{irr}^{'}$(0.5) can be expressed as,
\begin{equation}
A_{irr}^{'}(0.5)\,=\,
\left\{
\begin{array}{l}
A_{irr}-A_{irr}^{'}(0.0)\qquad\,\theta\ge i\\
A_{irr}\qquad\qquad\qquad\quad\theta<i\\
\end{array}
\right.
\end{equation}
By using Equation A2, the flux ratio between the phases 0.5 and 0.0 (i.e., F$_{irr}=f_{0.5}/f_{0.0}$) can be calculated for a given $\theta$. Based on the normalized area of the irradiation region, S$_{irr}$=A$_{irr}$/A$_{star}$ (i.e., $\cos{\theta}=1\,-\,2S_{irr}$), F$_{irr}$ can be described as,
\begin{large}
\begin{equation}
F_{irr}\,=\,
\left\{
\begin{array}{l}
\frac{(\pi-2\pi S_{irr}-C_{irr})\,T_{star}^{4}+(2\pi S_{irr}-C_{irr})\,T_{irr}^{4}}{(\pi-C_{irr})\,T_{star}^{4}+C_{irr}\,T_{irr}^{4}}\qquad\quad\;\theta\ge i\\
\\
(1-2S_{irr})[\frac{T_{irr}}{T_{star}}]^{4}-2(1+S_{irr})\qquad\quad\;\theta<i\\
\end{array}
\right.
\end{equation}
\end{large}
The relations of F$_{irr}$ vs. S$_{irr}$ for the three DN and one pre-CV system discussed here are plotted in Fig. 5.

\begin{figure}\figurenum{A}
\centering
\includegraphics[width=10.0cm]{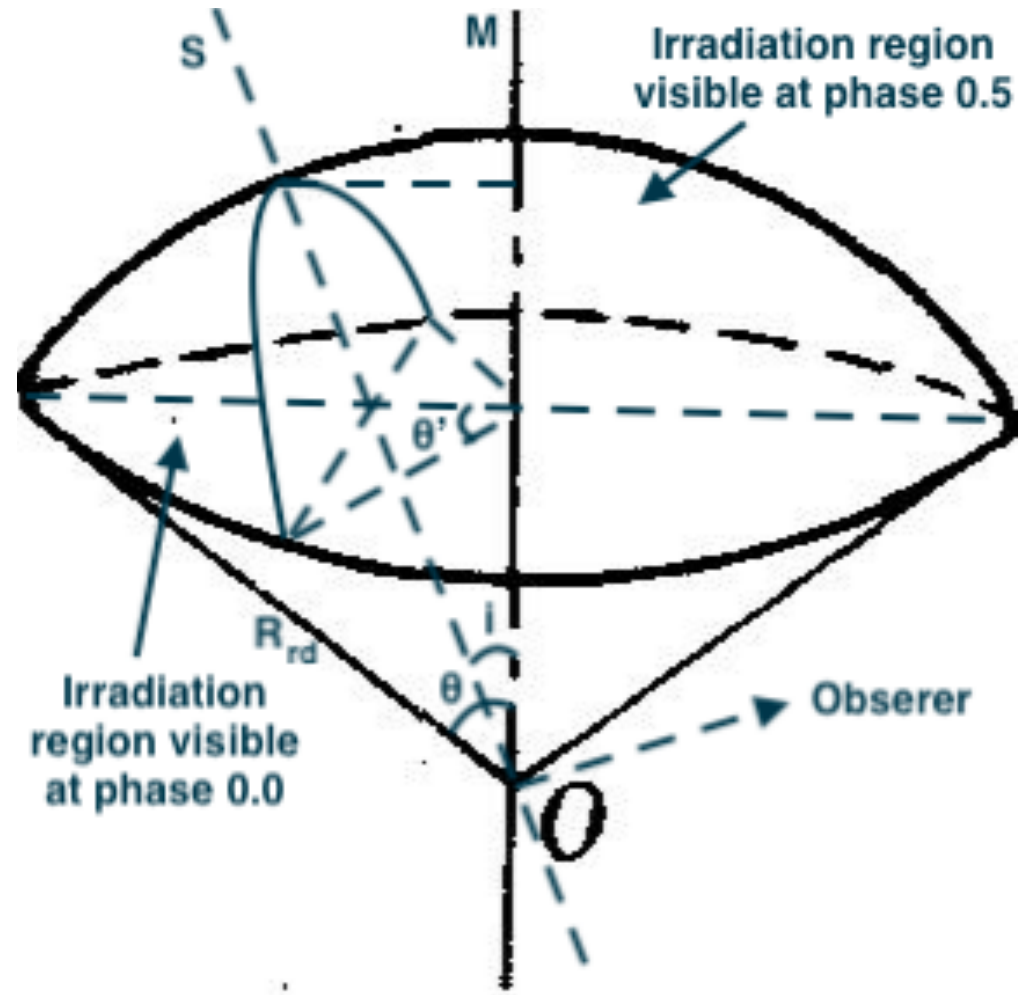}
\caption{\small{This sketch displays an irradiation region of the secondary star with a segment shape. The center of the secondary is at O point. The OM line is located on the orbital plane. While the OS line indicates a plane vertical to the line of sight. $\theta$ is the half of the opening angle of the irradiation region.}}
\end{figure}

\clearpage

\section{A new observed optical spectrum of WD\,1144+011}

A 1.8 arcsec slit and the G4 grating with a low resolution of 1953 (i.e., 2.97\,\AA pixel$^{-1}$) was used with an exposure time of 1,800\,sec. Flux standards and FeAr lamps were used along with IRAF \footnote{IRAF is distributed by the National Optical Astronomy Observatory, which is operated by the Association of Universities for Research in Astronomy (AURA) under cooperative agreement with the National Science Foundation.} reductions to produce the final calibrated spectrum shown in Fig. B. 

This spectrum is dominated by a strong and fairly broad H$_{\alpha}$ emission line and several molecular absorption bands (e.g., MgH5211\AA\,and CaH6358\AA), and is similar to the previous spectrum obtained by \cite{ber92}. Thus, our spectrum supports the previous spectral classification of WD\,1144+011 derived by \citep{ber92}. An active M star can show variable H$_{\alpha}$ in emission. However, our spectrum shows a flatter blue continuum with a higher red flux level. The measured average equivalent width of H$_{\alpha}$ is close to 100\,\AA, although this line is difficult to measure accurately due to the uncertainty of the continuum. What is clear from both our spectrum and that of \cite{ber92} is that there is no prominent blue continuum from a hot white dwarf or disk and that the Balmer lines are in emission and of variable strength. The changes in flux levels between our spectrum and the spectrum presented in \cite{ber92} could be due to phase differences when the two spectra were taken.

\begin{figure}\figurenum{B}
\centering
\includegraphics[width=14.0cm]{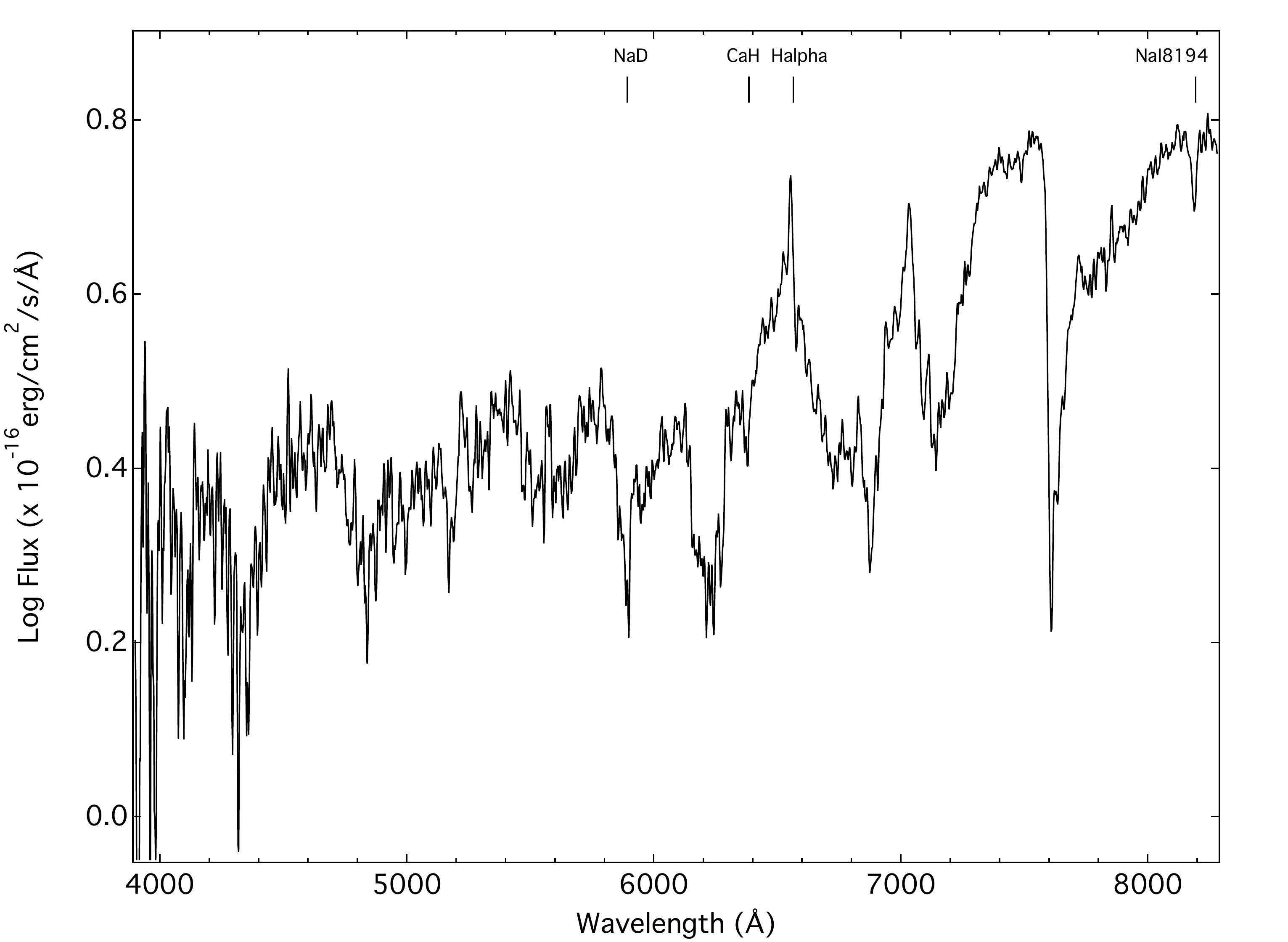}
\caption{\small{The optical spectrum of WD\,1144+011 covering a wavelength range from 3,900\AA\,to 8,300\AA\,was taken by using the Xinglong 2.16\,m telescope on January 22, 2017. The characteristic spectral lines are marked.}}
\end{figure}

\begin{table}
\caption{All parameters used to calculate light curves by using XRBinary.}
\begin{center}
\begin{tabular}{lccc}
\hline\hline
Parameters & Values & Statements\\
\hline
\textbf{Fixed} &&\\
\hline
STAR2TILES & 5,000 & tiles covering star surface\\
DISKTILES & 20,000 & tiles covering disk surface\\
BANDPASS (\AA) & 4,200$\sim$9,000 & bandpass of K2 light curve\\
H$_{pow}$ & 1.1 & power index of H$_{disk}$\\
L$_{wd}$ & -- & white dwarf luminosity\\ 
T$_{wd}$ & -- & white dwarf temperature\\
\tablenotemark{a} R$_{in}$ & -- & inner radius of disk\\
\hline
\textbf{Adjustable} &&\\
\hline
\multicolumn{2}{l}{\quad Components} &\\
\hline
q$_{orb} $ & -- & mass ratio\\
i$_{orb}$ & -- & orbital inclination\\
M$_{wd}$ & -- & white dwarf mass\\
T$_{rd}$ & -- & red dwarf temperature\\
R$_{out}$ & -- & outer radius of disk\\
H$_{edge}$ & -- & H$_{disk}$ at the outer edge\\
$\xi$ & -- & power index of T$_{disk}$\\
L$_{d0}$ & -- & disk luminosity\\
\hline
\multicolumn{2}{l}{\quad Hotspot at the edge of disk} &\\
\hline
T$_{es}$ & -- & temperature\\
$\zeta_{esmid}$ & -- & centering phase\\
$\zeta_{eswidth}$ & -- & full width\\
\hline
\multicolumn{2}{l}{\quad Hotspot on the surface of disk} &\\
\hline
$\zeta_{ssmin}$ & -- & lower limit of boundary in $\zeta$ direction\\
$\zeta_{ssmax}$ & -- & upper limit of boundary in $\zeta$ direction\\
R$_{ssmin}$ & -- & lower limit of boundary in radius direction\\
R$_{ssmax}$ & -- & uppr limit of boundary in radius direction\\
T$_{ratio}^{ss}$ & -- & fractional change in T$_{disk}$\\
\hline\hline
\end{tabular}
\end{center}
\tablenotetext{a}{\,R$_{in}$ can be adjustable for IPs due to the disrupted inner disk.}
\end{table}

\begin{table}
\caption{The models used in XRBinary.}
\begin{center}
\begin{tabular}{cccc}
\hline\hline
\tablenotemark{a} Component & Model-0 & Model-1 & Model-2\\
\hline
White dwarf & 1 & 1 & 1\\
Red dwarf & 1 & 1 & 1\\
Accretion disk & 0 & 1 & 1\\
Hotspot on the disk surface & 0 & 0 & 1\\
\hline\hline
\end{tabular}
\end{center}
\tablenotetext{a}{\,The component included in the model or not is indicated by "1" or "0", respectively.}
\end{table}

\begin{table}
\caption{The zero points of the relative flux contributions from different component}
\begin{center}
\begin{tabular}{ccccc}
\hline\hline
\tablenotemark{a} Components & J0632+2536 & RZ\,Leo & TW\,Vir & WD\,1144+011\\
\hline
\tablenotemark{b} Binary & 65.1 & 66.5 & 20.4 & 94.3\\
\tablenotemark{c} Starspot & -- & -- & -- & 1.1\\
\tablenotemark{d} Disk & 11.7 & 0.0 & 63.0 & --\\
Hotspot$^{es}$ & 0.0 & 0.0 & 0.0 & --\\
Hotspot$^{ss}$ & -- & 2.6 & 3.4 & --\\
\hline\hline
\end{tabular}
\end{center}
\tablenotetext{a}{\,The relative flux contributions are in percentage.}
\tablenotetext{b}{\,Only consist of white dwarf and red dwarf.}
\tablenotetext{c}{\,Starspots on the red dwarf.}
\tablenotetext{d}{\,Accretion disk without hotspot.}
\end{table}

\begin{table}
\caption{The results of four systems estimated from the Gaia database.}
\begin{center}
\begin{tabular}{cccccccc}
\hline\hline
Name & T$_{eff}$ & Parallax & D$_{g}$ & L$_{all}$ & \tablenotemark{a} BC$_{v}$ & \tablenotemark{b} CKp$_{2}$ & \tablenotemark{c} Kp$_{2}$\\
\hline
& K & mas & pc & $\times10^{32}\,erg\,s^{-1}$ && mag & mag\\
\hline
J0632+2536 & 5255.5 & 2.24(5) & 446($\pm$10) & 13.2 & -0.53(4) & 14.67(7) & 14.4\\
RZ Leo & -- & 3.6(3) & 278($\pm$23) & 0.2 & -1.4(4) & 19.1(4) & 17.6\\
TW Vir & 4850 & 2.3(1) & 435($\pm$19) & 5.8 & -1.1(1) & 16.0(2) & 15.9\\
WD1144+011 & 3657.5 & 4.7(1) & 213($\pm$5) & 4.8 & -1.8(7) & 15.5(7) & 16.0\\
\hline\hline
\end{tabular}
\end{center}
\tablenotetext{a}{\,Interpolated bolometric corrections based on BC$_{v}$ tabulations by \cite{cas14}.}
\tablenotetext{b}{\,Calculated K2 magnitude by using the distance and L$_{all}$.}
\tablenotetext{c}{\,Observed K2 magnitude corresponding to L$_{all}$.}
\end{table}

\begin{table}
\caption{Photometric solutions for three DN.}
\begin{center}
\begin{tabular}{lccc}
\hline\hline
\tablenotemark{a} Parameter & J0632+2536 & RZ\,Leo & TW\,Vir\\
\hline
\tablenotemark{b} Model type & Model-1 & Model-2 & Model-2$^{irr}$\\
\hline
\textbf{Orbit} &&&\\
\hline
q$_{orb} (M_{rd}/M_{wd})$ & 0.8(1) & 0.080(6) & 0.41(5)\\
i$_{orb}$ (degree) & 60.3(5) & 61.0(9) & 44.3(5)\\
\hline
\textbf{White dwarf} &&&\\
\hline
M$_{wd} (M_{\odot})$ & 0.81(8) & 1.26(3) & 1.10(3)\\
\tablenotemark{c} R$_{wd} (R_{\odot})$ & 0.01 & 0.005 & 0.007\\
\tablenotemark{d} T$_{wd}$ & 25.4 & 15.0 & 19.1\\
\tablenotemark{c} L$_{wd}$ & 1.5 & 0.045 & 0.23\\
\hline
\textbf{Red dwarf} &&&\\
\hline
M$_{rd} (M_{\odot})$ & 0.7(1) & 0.101(8) & 0.45(6)\\
\tablenotemark{c} R$_{rd} (R_{\odot})$ & 0.81 & 0.16 & 0.48\\
T$_{rd}$ & 4.54(8) & 3.71(3) & 4.00(4)\\
\tablenotemark{c} L$_{rd}$ & 9.6 & 0.17 & 2.0\\
\hline
\textbf{Accretion disk} &&&\\
\hline
R$_{in} (R_{\odot})$ & \tablenotemark{d} 0.01 & 0.211(2) & \tablenotemark{d} 0.007\\
R$_{out} (R_{\odot})$ & 1.22(1) & 0.401(2) & 0.536(3)\\
H$_{edge} (R_{\odot})$ & 0.091(4) & 0.0801(6) & 0.1174(3)\\
$\xi$ & \tablenotemark{e} -0.11 & -0.12(6) & -0.11(4)\\
L$_{d0}$ & 3.9(3) & 0.00061(6) & 3.36(6)\\
\tablenotemark{c} L$_{d}$ & 4.1 & 0.023 & 4.0\\
\hline
\multicolumn{2}{l}{\quad Hotspot at the edge of the disk} &&\\
\hline
T$_{es}$ & 6.2(1) & 3.71(2) & 4.80(2)\\
$\zeta_{esmid}$ (phase) & 0.844(2) & 0.723(1) & 0.727(6)\\
$\zeta_{eswidth}$ (phase) & 0.033(2) & 0.057(4) & 0.296(3)\\
\hline
\multicolumn{2}{l}{\quad Hotspot on the surface of disk} &&\\
\hline
$\zeta_{ssmin}$ (phase) & -- & 0.38(2) & 0.32(5)\\
$\zeta_{ssmax}$ (phase) & -- & 0.56(2) & 0.74(3)\\
R$_{ssmin} (R_{\odot})$ & -- & \tablenotemark{e} 0.211 & 0.489(8)\\
R$_{ssmax} (R_{\odot})$ & -- & 0.217(5) & \tablenotemark{e} 0.536\\
T$_{ratio}^{ss}$ & -- & 7.2(2) & 1.20(2)\\
\hline
$\chi^{2}$ & 11.9 & 2.3 & 0.66\\
\hline\hline
\end{tabular}
\end{center}
\tablenotetext{a}{\,The unit of temperature and luminosity is 10$^{3}\,$K and 10$^{32}\,erg\,s^{-1}$, respectively.}
\tablenotetext{b}{\,Model with a superscript of ``irr" denotes that the irradiation effect is included in XRBinary.}
\tablenotetext{c}{\,Calculated by XRBinary.}
\tablenotetext{d}{\,Fixed in NMfit program.}
\tablenotetext{e}{\,Insensitive to the observed light curves.}
\end{table}

\begin{table}
\caption{The estimated parameters of the secondaries of four systems}
\begin{center}
\begin{tabular}{ccccc}
\hline\hline
\tablenotemark{a} Name & $\dot{M}_{rd}$ & $\tau_{\dot{M}}$ & $\tau_{kh}$ & $\tau_{\dot{M}}/\tau_{kh}$\\
\hline
& $10^{-10}M_{\odot}$/yr & $10^{9}$yr & $10^{9}$yr &\\
\hline
J0632+2536 & 2.5($\pm$0.2) & 2.8($\pm$0.5) & 0.076 & 37\\
RZ\,Leo & 0.0311($\pm$0.0008) & 32.4($\pm$2.7) & 0.45 & 72\\
TW\,Vir & 1.92($\pm$0.05) & 2.3($\pm$0.3) & 0.25 & 9\\
\tablenotemark{b} WD\,1144+011 & 67.8($\pm$16.0) & 0.1($\pm$0.03) & 0.006 & 18\\
\tablenotemark{c} WD\,1144+011 & 7.9($\pm$1.8) & 1.0($\pm$0.3) & 0.027 & 37\\
\hline\hline
\end{tabular}
\end{center}
\tablenotetext{a}{\,Calculations by using the parameters listed in Tables 5 and 7.}
\tablenotetext{b}{\,For the model-1.}
\tablenotetext{c}{\,For the model-2.}
\end{table}

\begin{table}
\caption{Three photometric solutions for the pre-CV WD\,1144+011.}
\begin{center}
\begin{tabular}{lccc}
\hline\hline
\tablenotemark{a} Parameters & \multicolumn{3}{c}{WD\,1144+011}\\
\hline
\tablenotemark{b} Model type & Model-0$^{irr}$ & Model-1 & Model-2\\
\hline
\textbf{Orbit} &&&\\
\hline
q$_{orb} (M_{rd}/M_{wd})$ & 0.49(2) & 0.99(2) & 0.92(9)\\
i$_{orb}$ (degree) & 14.3(4) & 17.7(4) & 26.3(6)\\
\hline
\textbf{White dwarf} &&&\\
\hline
M$_{wd} (M_{\odot})$ & 1.11(8) & 0.68(16) & 0.9(2)\\
\tablenotemark{c} R$_{wd} (R_{\odot})$ & 0.007 & 0.012 & 0.009\\
\tablenotemark{d} T$_{wd}$ & 27.1 & 27.0 & 25.7\\
\tablenotemark{c} L$_{wd}$ & 0.91 & 2.5 & 1.3\\
\hline
\textbf{Red dwarf} &&&\\
\hline
M$_{rd} (M_{\odot})$ & 0.54(4) & 0.67(16) & 0.8(2)\\
\tablenotemark{c} R$_{rd} (R_{\odot})$ & 0.87 & 0.97 & 1.03\\
T$_{rd}$ & 3.5(1) & 7.44(7) & 5.23(6)\\
\tablenotemark{c} L$_{rd}$ & 4.1 & 100.7 & 27.7\\
\hline
\multicolumn{2}{l}{\quad Bright starspot on the red dwarf} &&\\
\hline
$\theta^{sp}$ (degree) & 38(3) & -- & --\\
$\phi^{sp}$ (degree) & \tablenotemark{e} 0.2 & -- & --\\
R$^{sp}$ (degree) & 27.7(7) & -- & --\\
T$_{ratio}^{sp}$ & 1.097(3) & -- & --\\
\hline
\textbf{Accretion disk} &&&\\
\hline
R$_{in} (R_{\odot})$ & -- & \tablenotemark{d} 0.01 & \tablenotemark{d} 0.009\\
R$_{out} (R_{\odot})$ & -- & 0.597(5) & 0.808(8)\\
H$_{edge} (R_{\odot})$ & -- & 0.363(8) & 0.176(3)\\
$\xi$ & -- & \tablenotemark{e} -0.11 & -0.18(4)\\
L$_{d0}$ & -- & 13.7($\pm$2.7) & 13.5($\pm$2.3)\\
\tablenotemark{c} L$_{d}$ & -- & 23.0 & 19.4\\
\hline
\multicolumn{2}{l}{\quad Hotspot at the edge of the disk} &&\\
\hline
T$_{es}$ & -- & 6.50(5) & 5.38(7)\\
$\zeta_{esmid}$ (phase) & -- & 0.623(4) & 0.548(8)\\
$\zeta_{eswidth}$ (phase) & -- & 0.41(2) & 0.17(1)\\
\hline
\multicolumn{2}{l}{\quad Hotspot on the surface of disk} &&\\
\hline
$\zeta_{ssmin}$ (phase) & -- & -- & 0.129(6)\\
$\zeta_{ssmax}$ (phase) & -- & -- & 0.221(2)\\
R$_{ssmin} (R_{\odot})$ & -- & -- & 0.605(1)\\
R$_{ssmax} (R_{\odot})$ & -- & -- & \tablenotemark{e} 0.808\\
T$_{ratio}^{ss}$ & -- & -- & 1.90(4)\\
\hline
$\chi^{2}$ & 8.6 & 7.9 & 8.8\\
\hline\hline
\end{tabular}
\end{center}
\tablenotetext{}{All footnotes are the same as those of Table 5.}
\end{table}

\clearpage

\begin{figure}
\centering
\includegraphics[width=14.0cm]{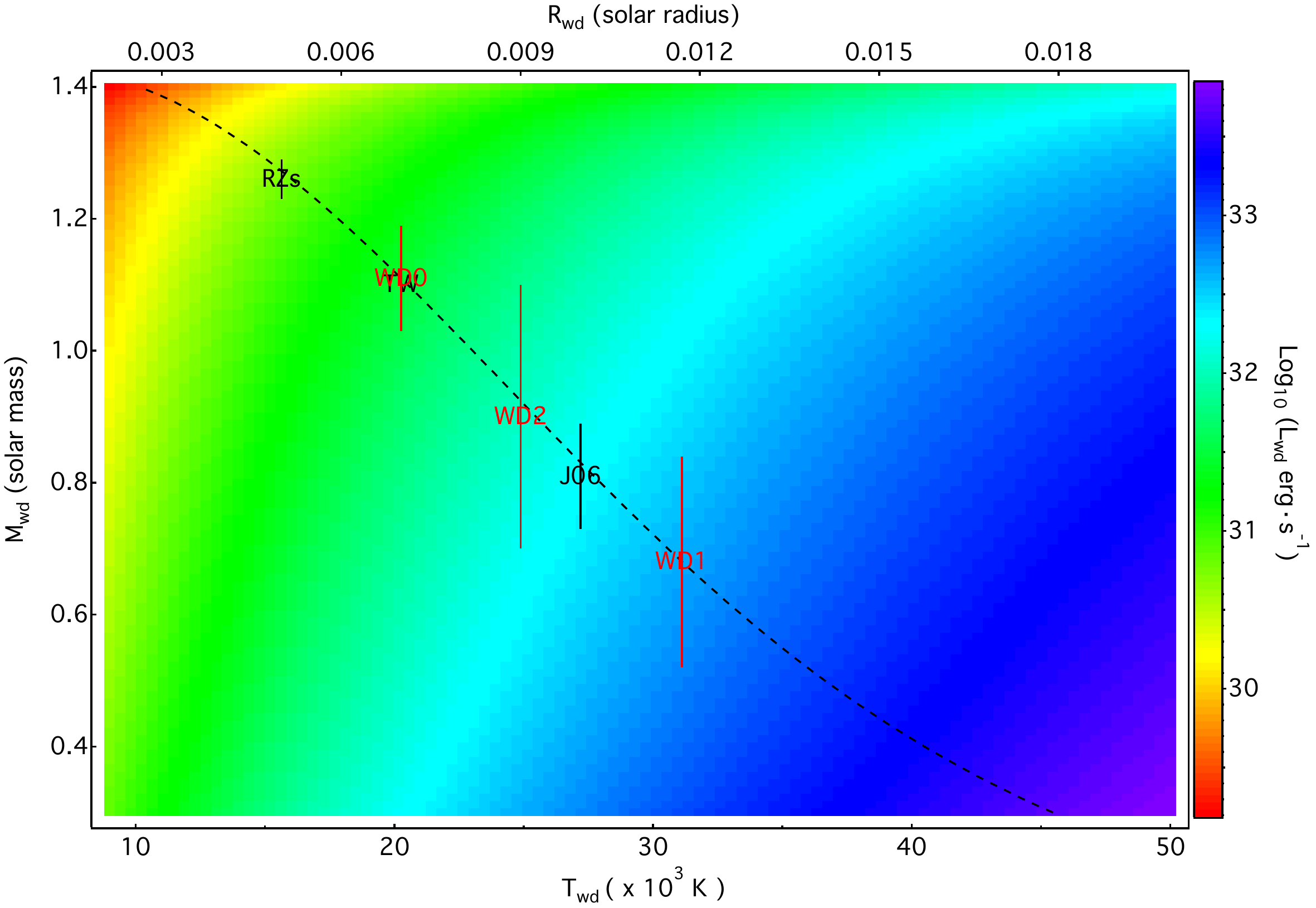}
\caption{\small{The colormap of the luminosity of the primary white dwarf corresponding to M$_{wd}$ and T$_{wd}$. The dash line represents the typical mass-radius relationship of the white dwarf. The datapoint marked by RZs refer to the model of RZ\,Leo derived from the light curve with the phase zero at the secondary dip. The three datapoints marked by WD0, WD1 and WD2 with the red color refer to the models of WD\,1144+011 by using the revised model-0, model-1 and model-2, respectively. The left two datapoints marked by TW and J06 refer to the models of TW\,Vir and J0632+2536, respectively.}}\label{Figure 1}
\end{figure}

\begin{figure}
\centering
\includegraphics[width=14.0cm]{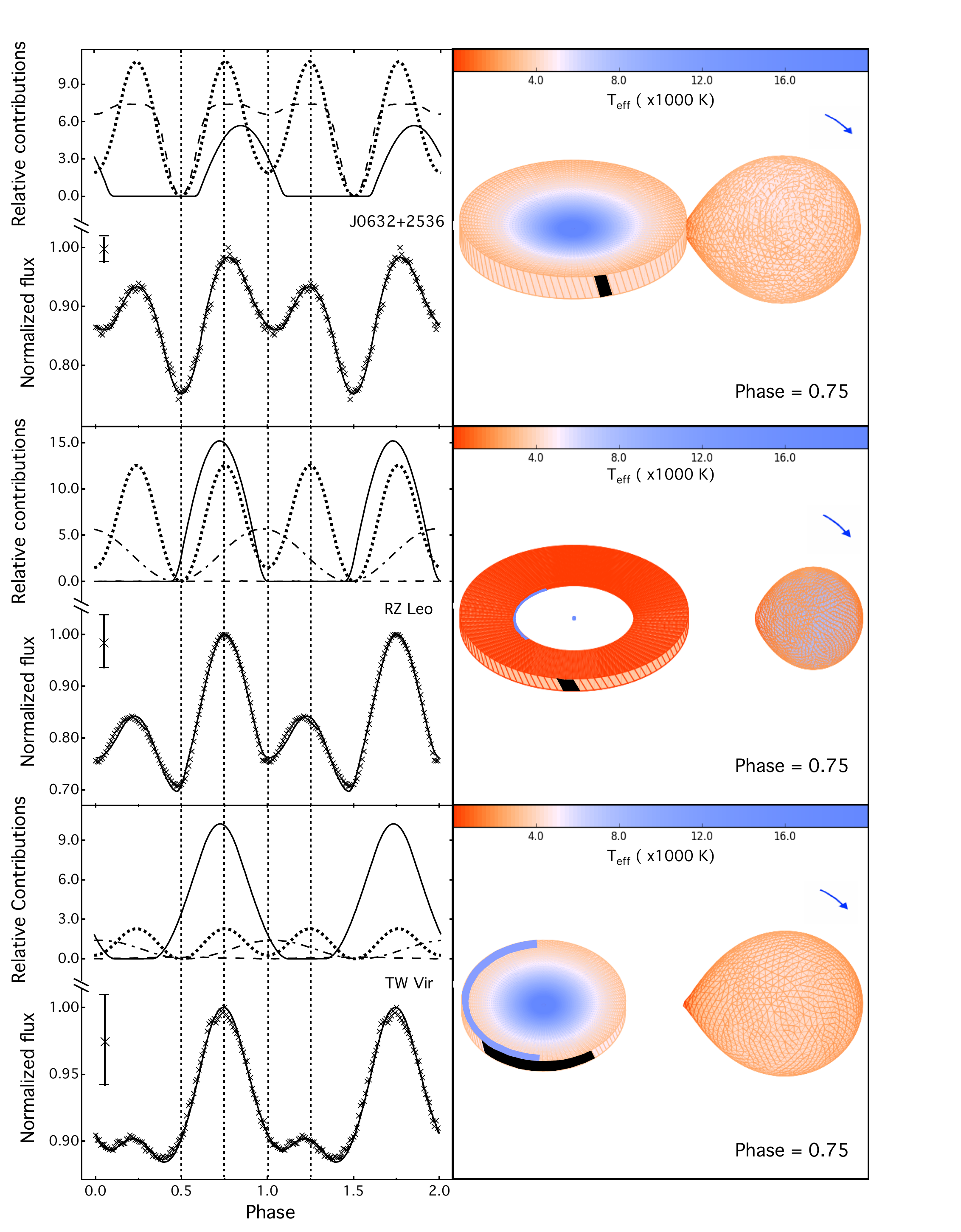}
\caption{\small{The phased and binned light curves of J0632+2536, RZ\,Leo and TW\,Vir superimposed with their best-fitting light curves are plotted in the left three panels from top to bottom, respectively. The flux of all light curves are normalized. The relative flux contributions from different model component for three systems are plotted in unit of percentage. The dot and dash lines refer to the contributions from two stellar components (white dwarf and red dwarf) and the accretion disk, respectively. The solid and dash dot lines denote the contributions from the hotspots at the edge of the disk and on the surface of the disk, respectively. They indicate what component is actually contributing to the actual light curve. The right three panels show their corresponding 2D binary configurations at phase 0.75 by using the Phoebe 2.0. The color denotes the effective temperature. In order to visualize the hotspot at the edge of the disk, they are filled with black rather than the color picked from the color bar, since the small temperature difference between the hotspot and the neighboring region of the disk can seriously reduce the contrast of the hotspot. The arrow denotes a clockwise rotation direction of the binary system.}}\label{Figure 2}
\end{figure}

\begin{figure}
\centering
\includegraphics[width=16.0cm]{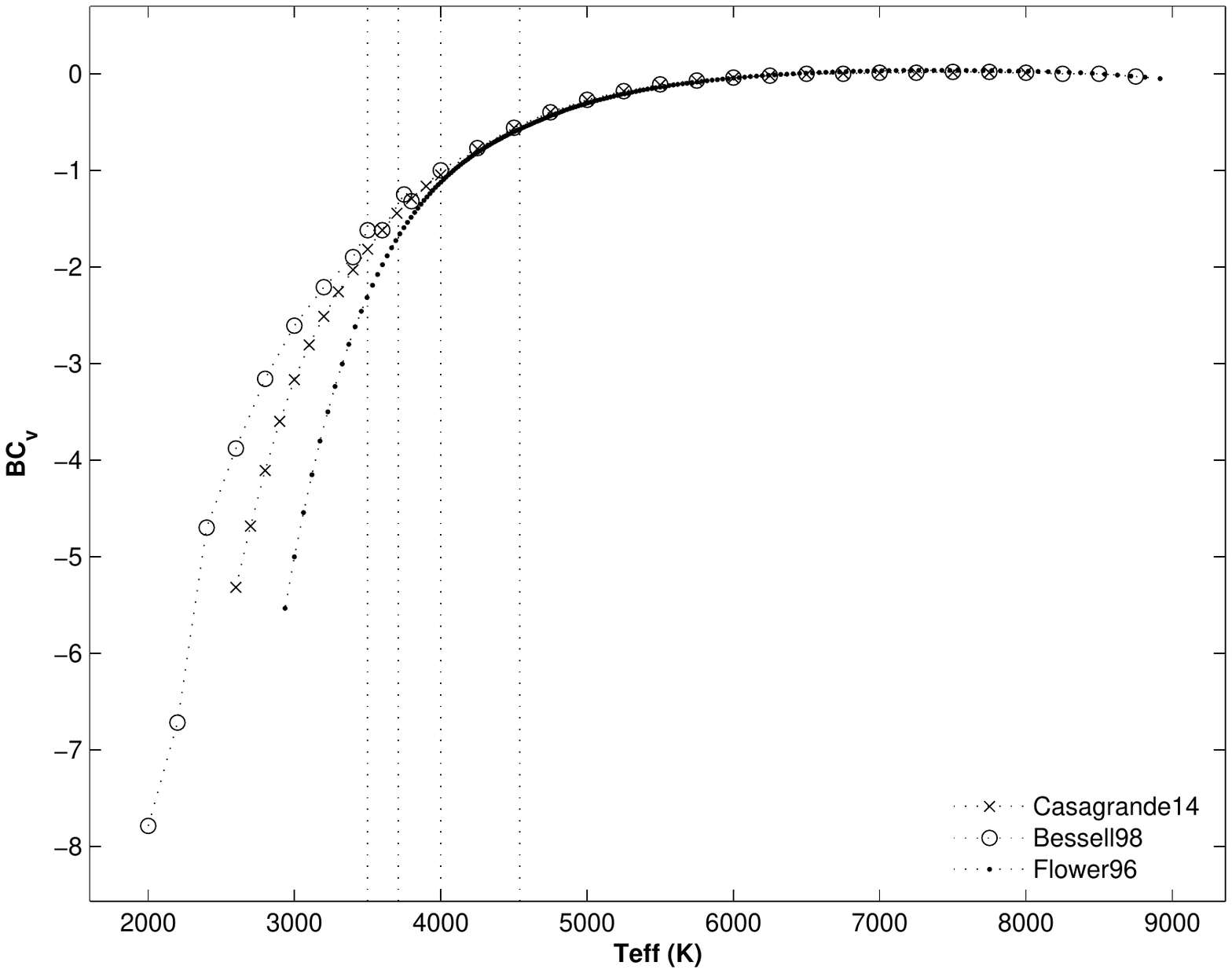}
\caption{\small{The diagram of bolometric correction in V band (BC$_{v}$) against the effective temperature of a main sequence star (T$_{eff}$). Three legends of Casagrande14, Bessell98 and Flower96 refer to the BC$_{v}$ tabulations given by \cite{cas14}, \cite{bes98} and \cite{flo96},respectively. Four dot lines from left to right denote T$_{rd}$ of WD\,1144+011, RZ\,Leo, TW\,Vir and J0632+2536, respectively.}}\label{Figure 3}
\end{figure}

\begin{figure}
\centering
\includegraphics[width=14.0cm]{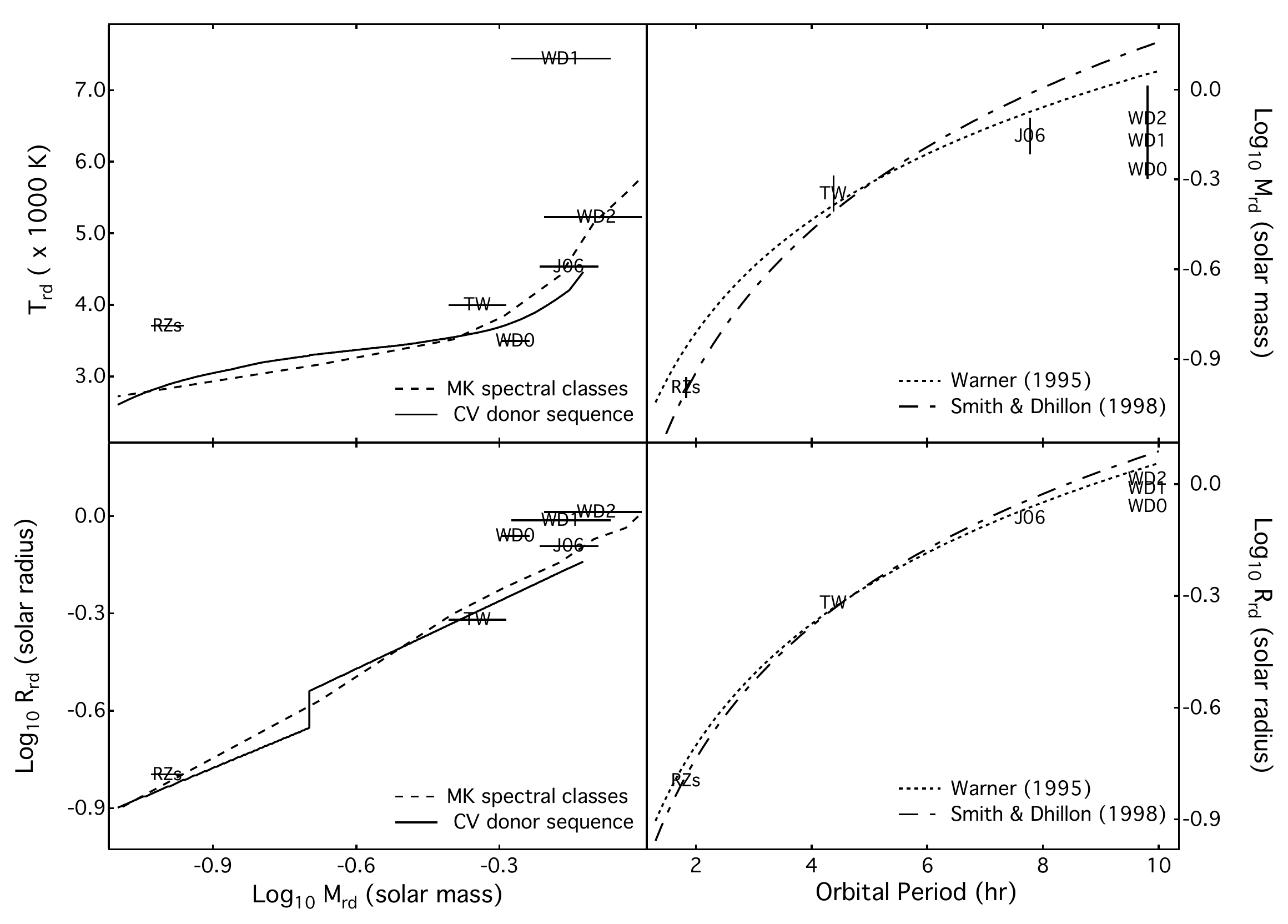}
\caption{\small{Four relationships of the secondaries. Top left panel: the relationship of the logarithm of mass and the effective temperature. Bottom left panel: the logarithm of mass-radius relationship. Top right panel: the period-mass relationship. Bottom right panel: the period-radius relationship. The dash and solid lines denote the relationships based on the isolated low-mass stars \citep{cox00}, and the semi-empirical CV donor sequence \citep{kni06,kni11}, respectively. The dash-dot and dot lines describe the relationships derived by \cite{smi98} and \cite{war03}, respectively. The symbols of the datapoints are the same as those used in Fig. 1.}}\label{Figure 4}
\end{figure}

\begin{figure}
\centering
\includegraphics[width=16.0cm]{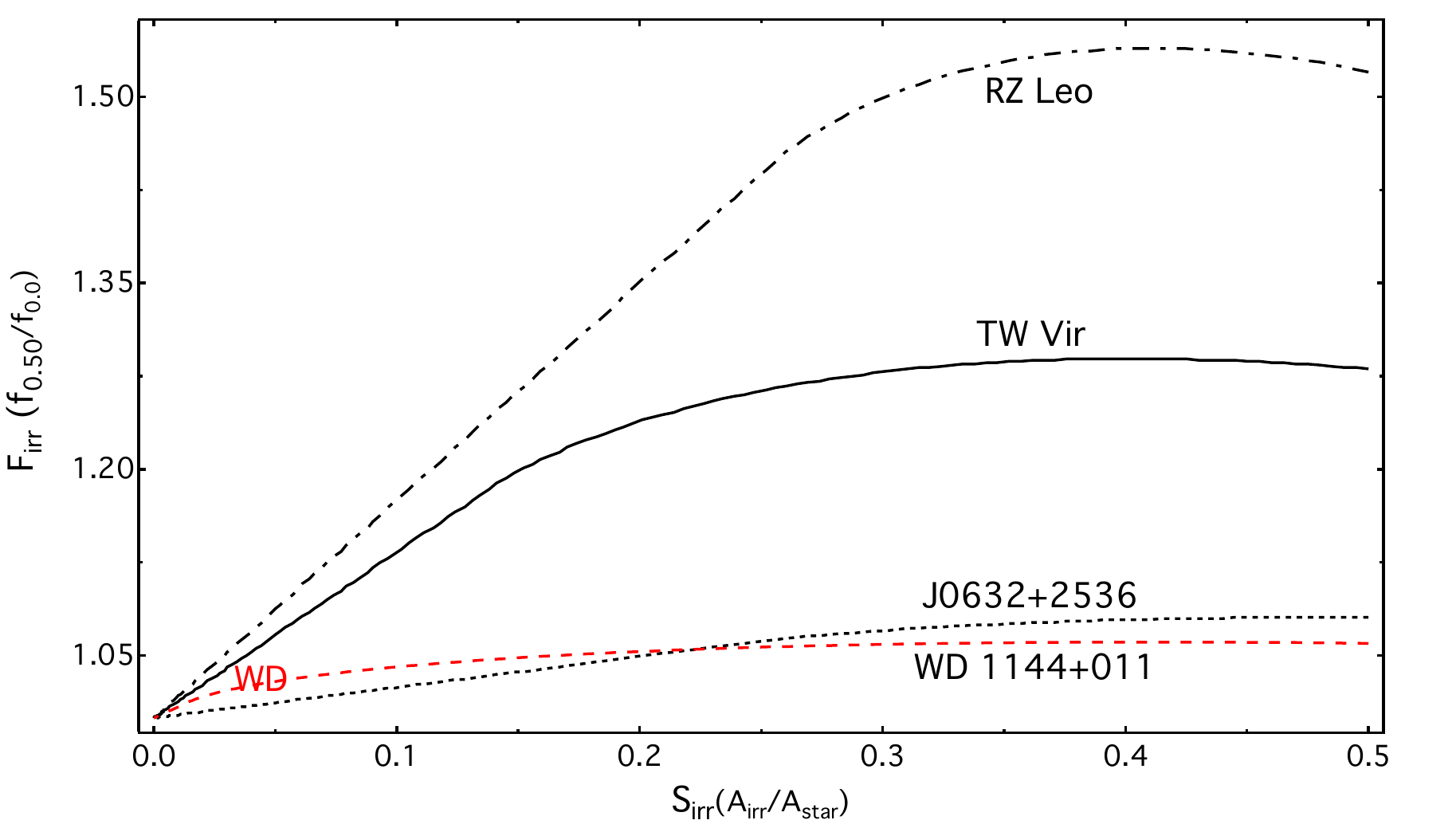}
\caption{\small{The diagram of the flux ratio between phases 0.5 and 0.0, F$_{irr}$, changing along with a normalized area of the irradiation region S$_{irr}$. The red symbol WD denotes S$_{irr}\sim0.05$ corresponding to F$_{obs}=1.03$ estimated from the single-hump modulation of WD\,1144+011.}}\label{Figure 5}
\end{figure}

\begin{figure}
\centering
\includegraphics[width=16.0cm]{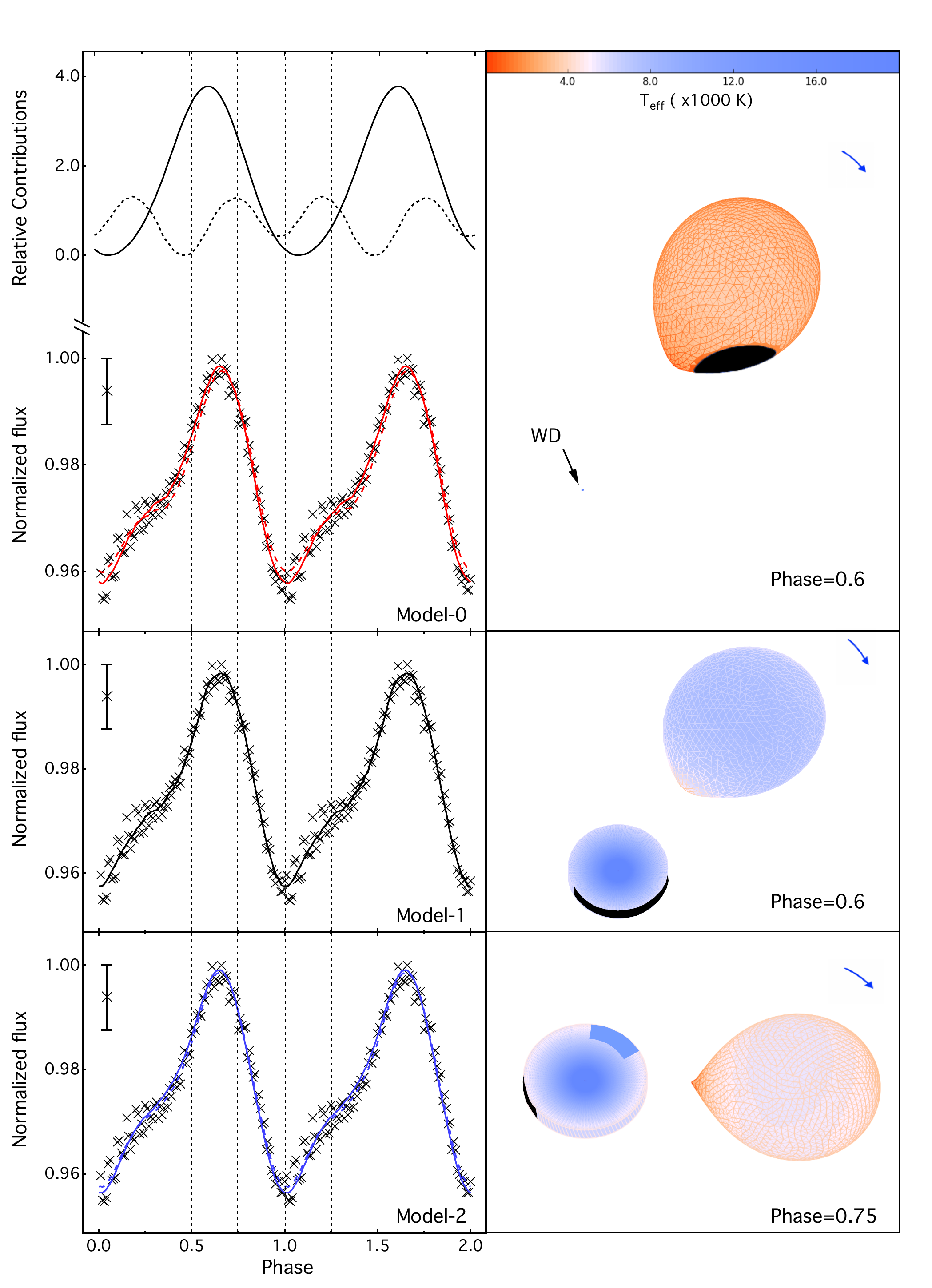}
\caption{\small{The phased and binned light curve of WD\,1144+011 superimposed with the best-fitting light curves denoting with the red, black and blue lines calculated by using the model-0, model-1 and model-2 are plotted in the top, middle and bottom left panels, respectively. The solid and dash lines respectively refer to two types of calculations including and excluding the irradiation effect in XRBinary. The right three panels from the top to bottom show the corresponding 2D binary configurations at phases 0.6, 0.6 and 0.75, respectively. All symbols are the same as that of Fig. 2.}}\label{Figure 6}
\end{figure}

\end{CJK*}
\end{document}